\documentclass[a4paper,12pt]{article}
\usepackage{jheppub}  

\usepackage{graphicx}
\usepackage{color}
\usepackage[table]{xcolor}
\usepackage{slashed}
\usepackage{multirow}
\usepackage{float}
\usepackage{afterpage}
\usepackage{lineno}
\usepackage{orcidlink}
\usepackage{xspace}

\newcommand{\dsum}{\displaystyle\sum}

\newcommand{\sumab}{\dsum_{a>b}}

\newcommand{\ith}{i^\text{th}}
\newcommand{\jth}{j^\text{th}}
\newcommand{\kt}{\ensuremath{k_t}\xspace}

\newcommand{\dij}{d_{ij}}
\newcommand{\diB}{d_{iB}}

\newcommand{\pt}{p_T}
\newcommand{\pti}{{p_T}_i}
\newcommand{\ptj}{{p_T}_j}\newcommand{\ptk}{{p_T}_k}
\newcommand{\pta}{{p_T}_a}
\newcommand{\ptb}{{p_T}_b}
\newcommand{\ptc}{{p_T}_c}
\newcommand{\ptJ}{{p_T}_J}

\newcommand{\dR}{\Delta R}
\newcommand{\dRij}{\Delta R_{ij}}
\newcommand{\dRab}{\Delta R_{ab}}
\newcommand{\dRJa}{\Delta R_{Ja}}

\newcommand{\sigmaRi}{\sigma_i}
\newcommand{\sigmaRj}{\sigma_j}

\newcommand{\Rd}{R_d}
\newcommand{\Rdi}{{R_d}_i}

\newcommand{\Ravgi}{\rho_i}
\newcommand{\Ravgj}{\rho_j}

\newcommand{\Rnot}{R_0}

\newcommand{\Vj}{{\bf Vj}\xspace}
\newcommand{\tj}{{\bf tj}\xspace}
\newcommand{\jj}{{\bf jj}\xspace}
\newcommand{\jjth}{{\bf jj300}\xspace}
\newcommand{\jjfh}{{\bf jj500}\xspace}
\newcommand{\Vjth}{{\bf Vj300}\xspace}
\newcommand{\Vjfh}{{\bf Vj500}\xspace}
\newcommand{\tjfh}{{\bf tj500}\xspace}

\newcolumntype{L}[1]{>{\raggedright\let\newline\\\arraybackslash}p{#1}}

\newcolumntype{C}[1]{>{\centering\arraybackslash}p{#1}}

\newcolumntype{R}[1]{>{\raggedleft\arraybackslash}p{#1}}

\begin{document}
\preprint{}

\title{Generalised Dynamic Radius Jets for Robust Collider Analyses}

\author[a]{Songshaptak De\,\orcidlink{0000-0003-3174-7425}\,}
\emailAdd{songshaptak.de@ijs.si}
\emailAdd{deswaptak@gmail.com}
\affiliation[a]{Jo\v{z}ef Stefan Institute, Jamova 39, 1000 Ljubljana, Slovenia.}

\author[b]{\!\!, Tousik Samui\,\orcidlink{0000-0002-1485-6155}\,}
\emailAdd{tousiks@imsc.res.in}
\emailAdd{tousiksamui@gmail.com}
\affiliation[b]{The Institute of Mathematical Sciences, IV Cross Road, CIT Campus, Taramani, Chennai 600\,113, India.}

\author[c]{\!\!, and Ritesh K. Singh\,\orcidlink{0000-0001-7838-6191}\,}
\emailAdd{ritesh.singh@iiserkol.ac.in}
\affiliation[c]{Department of Physical Sciences, Indian Institute of Science Education and Research Kolkata, Mohanpur, 741\,246, India.}

\abstract{
Jets and their reconstructions play a central role in precision measurements and searches for new physics at hadron colliders. Conventional jet clustering algorithms employ a fixed radius parameter, which may not optimally describe events containing jets of varying characteristic sizes. Building on the recently proposed dynamic radius jet clustering framework, we construct several substructure-inspired dynamic radius prescriptions based on jet angularities and energy correlation functions. A detailed study of these algorithms is performed, including detector effects within the Delphes framework and in the presence of high pileup corresponding to an average of 150 interactions per event, with pileup contamination mitigated using the PUPPI algorithm. The performance of the proposed algorithms is compared with that of the standard anti-$k_t$ algorithm in boosted $Vj$ ($V=W^\pm,Z$) and $tj$ events against dijet backgrounds. Using jet substructure observables and multivariate analysis based on boosted decision trees, we find that the dynamic radius algorithms lead to improved reconstruction of boosted heavy-particle jets and achieve better signal-background discrimination compared to the conventional fixed radius anti-$k_t$ clustering.
}

\maketitle
\section{Introduction}
High-energy colliders, such as the Large Hadron Collider (LHC) at CERN, have played an indispensable role in modern particle physics experiments. These colliders not only confirmed theoretical predictions with great precision, but also made landmark discoveries through the direct production of heavy particles. With these discoveries, including the $W$ and $Z$ bosons at the Super Proton-Antiproton Synchrotron\,\cite{UA1:1983crd,UA2:1983tsx,UA1:1983mne,UA2:1983mlz}, the top quark at the Tevatron\,\cite{CDF:1995wbb,D0:1995jca}, and finally the Higgs boson at the Large Hadron Collider (LHC)\,\cite{ATLAS:2012yve,CMS:2012qbp}, the Standard Model (SM) of particle physics has been established as one of the most successful frameworks to describe elementary particles and their interactions seen in nature. With these successes, one direction of operating colliders is the luminosity frontier, such as the High-Luminosity LHC (HL-LHC)~\cite{Apollinari:2015wtw,Dainese:2019rgk,ZurbanoFernandez:2020cco,ATLAS:2025eii}, primarily aiming at precision measurements of the properties of known particles and their interactions. On the other hand, the energy frontier continues to push the energy of the machines further, particularly with the proposed future colliders~\cite{FCC:2018byv,FCC:2018evy,FCC:2018vvp}, which promise to explore physics at unprecedented energy scales. 

One of the striking phenomena at hadron colliders is the emergence of jets~\cite{Sterman:1977wj,UA1:1983hhd,Salam:2010nqg,Sapeta:2015gee}, collimated sprays of hadrons.
The coloured partons, such as quarks and gluons, once produced, undergo parton showering followed by hadronisation, resulting in jets. The identification and reconstruction of jets from the busy hadronic environment is an essential but challenging task. This is usually achieved through jet clustering algorithms~\cite{Bethke:1991wk,Brown:1991hx,Salam:2010nqg,Cacciari:2011ma}, which group the final state particles in jets based on their kinematic configurations.

In current high-energy colliders, not only are narrow jets originating from light quarks and gluons copiously produced, but heavy SM particles such as $W$, $Z$, Higgs bosons, and top quarks can also be produced with high transverse momenta\,\cite{Plehn:2009rk,Cui:2010km,Schatzel:2013wsr}. When such boosted heavy particles decay hadronically, their decay products become highly collimated due to the large Lorentz boost and often merge into a single jet, which carries their composite structure. These so-called fat jets or boosted jets exhibit internal structures that are characteristically different from ordinary narrow QCD jets\,\cite{Thaler:2008ju,Kaplan:2008ie,Abdesselam:2010pt,Altheimer:2013yza,Marzani:2019hun,Bonilla:2022wzp,De:2024puh}. While QCD jets are mostly one-pronged, fat jets tend to be multi-pronged due to the two- or three-body decays of their parent particles. Hence, the internal structure, called jet substructure (JSS), becomes a powerful tool to study and discriminate among various jet origins\,\cite{Chen:2013ola,Dasgupta:2013ihk,Larkoski:2017jix,Kogler:2018hem,Nayak:2019quy}.

Alongside the traditional narrow jets, the study of fat jets has become an essential component of collider physics\,\cite{CMS:2011xsa,Larkoski:2014zma,Freytsis:2014hpa,ATLAS:2014pkk,Adams:2015hiv,Brooijmans:2017klo,De:2020iwq,Dey:2021sug}. Understanding the substructure of these fat jets is not only possible but necessary in probing the boosted regime of SM particles and in searches for beyond the Standard Model (BSM) signals\,\cite{CMS:2020zge,Aguilar-Saavedra:2021rjk,CMS:2022suh,ATLAS:2022mlu,ATLAS:2023jdw,Sahu:2024fzi,CMS:2026cmh}. The structure of fat jets being different from narrow jets has given rise to a series of studies, both phenomenological and experimental\,\cite{Butterworth:2008iy,Chen:2011ah,CMS:2014hvu,CMS:2015fdn,Larkoski:2015kga,Chen:2015fca,ATLAS:2016bps,ATLAS:2019kwg,Maksimovic:2022fut,Kundu2024Jet}. Various JSS methods and observables, such as jet angularities\,\cite{Berger:2003iw,Almeida:2008yp,Larkoski:2014pca,Andersen:2016qtm}, energy correlation functions\,\cite{Larkoski:2013eya}, $N$-subjettiness\,\cite{Thaler:2010tr}, and Soft Drop\,\cite{Larkoski:2014wba}, have been proposed to distinguish fat jets from QCD backgrounds. These observables are found to be useful in studies related to tagging, grooming, and classification of jets, as well as in formulating search strategies at the colliders.

These JSS observables not only help in understanding the origin of jets, but also motivate the formation of jets. Particularly, they can be utilized to dynamically modify the jet radius parameter\,\cite{Krohn:2009zg,Mukhopadhyaya:2023rsb}. Given that jets arising from boosted decays tend to be wider than those from QCD, a static choice of radius may not always be optimal. Hence, a dynamical choice of jet radius, an approach in which the radius of each jet is modified depending on certain JSS features, provides an intuitive way of defining the radius parameters. 

In this work, we extend the idea of dynamic radius jet clustering\,\cite{Mukhopadhyaya:2023rsb}, in which the static jet-radius parameter of conventional clustering algorithms ($k_t$, anti-$k_t$, or CA) is replaced by a dynamical parameter. This dynamical parameter adapts the radius of a jet during the clustering process depending on the local kinematic structure of the evolving proto-jet. A specific realisation of this idea was previously proposed in Ref.\,\cite{Mukhopadhyaya:2023rsb}. In that, the central motivation was to determine the jet radius on a jet-by-jet basis using information encoded in the internal structure of the evolving proto-jet, rather than adopting a universal fixed radius parameter for each jet.

Building on our previous work, we generalise the dynamic radius framework beyond the particular choice of radius modifier used in that implementation\,\cite{Mukhopadhyaya:2023rsb}. Specifically, we exploit the connection between JSS and jet formation to construct a dynamic jet radius through substructure-inspired radius modifiers during the evolution of each jet. The jets originating from the boosted heavy particle decays are naturally assigned larger effective radii in order to capture the complete decay system, while ordinary QCD-like jets favour smaller radii. These characteristics are effectively captured by two important JSS observables: (i) jet angularities (JAs)\,\cite{Berger:2003iw,Almeida:2008yp,Larkoski:2014pca,Andersen:2016qtm} and (ii) energy correlation functions (ECFs)\,\cite{Larkoski:2013eya}. In this work, we use these observables to construct the dynamical radius of each jet. The details of the construction are presented in Sec.~\ref{sec:drAlgo}.

We then perform a detailed comparative study of the proposed dynamic radius constructions against conventional fixed radius jet clustering. The performance is evaluated using SM processes with final states $Vj$ ($V=W,Z$) and $tj$, considering $jj$ production as the background. In the boosted $Vj$ and $tj$ samples, jets of significantly different characteristic sizes are present, and their distinct signatures can be exploited through the dynamic radius construction. In contrast, dijet events typically contain jets of comparable size. To further investigate the properties of the reconstructed jets, we analyse several JSS observables, including the Soft Drop jet mass\,\cite{Larkoski:2014wba}, $N$-subjettiness\,\cite{Thaler:2010tr}, and ECF-based observables\,\cite{Larkoski:2013eya}. Using multivariate analysis techniques, we then demonstrate that incorporating JSS information into the determination of the jet radius improves the reconstruction of differently sized jets and enhances the separation between signal and background compared to conventional fixed radius clustering.

The rest of the paper is organised as follows. In Sec.~\ref{sec:jetAlgo}, we briefly review the conventional $k_t$-type jet clustering algorithms. In Sec.~\ref{sec:drAlgo}, we introduce the generalised framework of the dynamic radius algorithm. A brief discussion of the JSS observables employed in this work is presented in Sec.~\ref{sec:jssObs}. The analysis workflow is described in Sec.~\ref{sec:workflow}, while the results are presented in Sec.~\ref{sec:result}. Finally, we summarise our findings and present our outlook in Sec.~\ref{sec:summary}.

\section{Traditional Jet Clustering Algorithms: A Brief Review} \label{sec:jetAlgo}
The currently used algorithms at the LHC are the \kt-type sequential recombination algorithms\,\cite{Salam:2010nqg,Sapeta:2015gee} to capture both narrow QCD jets and boosted jets from heavy particles.
At the operational level, however, as far as the formation of jets is concerned, the same algorithm
with suitable choices of radius parameters is used for these two differently sized jets. For example, one uses a small radius of 0.4 for narrow QCD jets and a large radius of 0.8 or 1.0 for boosted fat jets. Therefore, these algorithms are inadequate in capturing the
differently sized jets appearing in a single event. 
The incorporation of varying radii for each jet is much needed for the current-day study.
Therefore, in order to capture differently sized jets in a single event as well as in a set of events, varying radii on a jet-by-jet basis would be a step forward.

We first explain key steps of the fixed radius sequential recombination algorithms, namely the \kt (KT)\,\cite{Catani:1991hj,Catani:1993hr,Ellis:1993tq}, anti-\kt (AK) \,\cite{Cacciari:2008gp}, or Cambridge-Aachen (CA)\,\cite{Dokshitzer:1997in,Wobisch:1998wt} algorithms. These algorithms start by assigning each particle or each proto-jet in an event a distance measure relative to every other particle or proto-jet and to the beam. The pairwise distances and beam distances are, respectively, defined as
\begin{eqnarray}
\dij &=& \min\left(\pti^{2n}, \ptj^{2n}\right)\,\dRij^2, \label{eqn:dij}\\
\diB &=& \pti^{2n} R^2, \label{eqn:diB}
\end{eqnarray} 
where $\pti$ is the transverse momentum of particle $i$, $\dRij$ is the separation between $\ith$ and $\jth$ particles in rapidity-azimuth ($y$-$\phi$) plane, and $R$ is the jet radius parameter. The exponent $n$ typically takes values of $-1$ for AK, $0$ for CA, and $1$ for KT algorithms. 
These algorithms then repeatedly (a) identify the smallest of all these distances; (b) if it is a pairwise distance, the two objects are recombined into a new proto-jet in the list, and the pair is removed from the list; (c) if it is a beam distance, the corresponding particle/proto-jet is declared as a final jet and removed from the list. The process continues until the list is exhausted. These algorithms are infrared and collinear (IRC) safe. In the case of the anti-\kt algorithm, the high-$\pt$ particles dominate the clustering because of the $\pt^{-2}$ weight in front of the $\Delta R$ distance. This leads to jets with nearly conical shapes, and it is one of the standard choices for LHC analyses.

The radius parameter in Eq.~(\ref{eqn:diB}) sets the typical cone sizes for the jets in an event, especially for the anti-\kt algorithm, where the jets are expected to be nearly conical. We therefore attempt to modify this fixed radius parameter to a dynamic one, which would dynamically evolve during the iterative procedure of the algorithm. So, our aim is to make $R \to \Rd$, a dynamical radius parameter, and it is set on a jet-by-jet basis. We further note that, for the fixed radius sequential recombination algorithm, the radius parameter $R$ can be placed in Eq.~(\ref{eqn:dij}) instead of Eq.~(\ref{eqn:diB}) without any alteration of the output of the clustering algorithm. In that case, the distance measures would look like
\begin{eqnarray}
\tilde\dij &=& \min\left(\pti^{2n}, \ptj^{2n}\right)\frac{\dRij^2}{R^2}, \label{eqn:dijtilde}\\
\tilde\diB &=& \pti^{2n}. \label{eqn:diBtilde}
\end{eqnarray}
Therefore, the choices of dynamising the radius parameter have two choices: (a) through Eqs.~(\ref{eqn:dij})--(\ref{eqn:diB}), or, alternatively, (b) through Eqs.~(\ref{eqn:dijtilde})--(\ref{eqn:diBtilde}). We will discuss these choices and various options in the next section. 

\section{Jet Substructure Observables}
\label{sec:jssObs}
In this section, we briefly review some of the key JSS observables that are used in our analysis.
\subsection{$N$-subjettiness}
\label{sec:nsubjettiness}
We first focus on the JSS methods that help probe the internal composition of jets. One such observable is $N$-subjettiness\,\cite{Thaler:2010tr,Thaler:2011gf}, which is designed to quantify how well a jet can be described in terms of $N$ hard ``subjets". This observable is particularly useful in the context of boosted objects such as $W$ bosons or top quarks, where the hadronic decay products can become collimated into a single fat jet, resulting in a two- or three-prong substructure.

For a given jet, $N$-subjettiness is defined as the minimum of the following measure
\begin{equation}
\tilde\tau_N^{(\xi)} = \frac{1}{d_0} \sum_k \ptk \min\left\{ (\Delta R_{1,k})^\xi, (\Delta R_{2,k})^\xi, \dots, (\Delta R_{N,k})^\xi \right\}, \label{eqn:nsubjettiness}
\end{equation}
where the sum runs over all constituents of the jet, $\ptk$ is the transverse momentum of constituent $k$, and $\Delta R_{J,k}$ is the distance in the $y$-$\phi$ plane between constituent $k$ and the $J$-th subjet axis. The minimization is performed over the $N$ subjet axes, and the corresponding minimum value defines $\tau_N$. The normalization factor is given by
\begin{equation}
d_0 = \sum_k \ptk R^\xi,
\end{equation}
where $R$ is the characteristic jet radius and $\xi$ is the angular exponent that controls the angular weighting of the radiation. The normalisation makes $\tau_N$ dimensionless, and its value ranges between 0 and 1. 

The more closely the constituents of a jet are aligned with the subjet axes, the smaller the value of $\tau_N$. For example, a jet originating from the hadronic decay of a boosted $W$ boson typically has a smaller $\tau_2$ than $\tau_1$,
whereas a jet originating from a boosted top quark tends to have 
a smaller $\tau_3$ than $\tau_2$. The ratios of the $N$-subjettiness variables have been shown to provide better discrimination between boosted hadronic objects and QCD jets\,\cite{Thaler:2011gf,CMS:2014rsx} than the individual $N$-subjettiness observables. In particular, we consider the following ratios in our analysis: \begin{equation}
\tau_{32} = \frac{\tau_3}{\tau_2} \qquad \mathrm{and} \quad
\tau_{21} = \frac{\tau_2}{\tau_1}.
\end{equation}
Since the same normalisation factor $d_0$ appears in the numerator and denominator, it cancels in these ratios.

In our analysis, we compute $\tau_1$, $\tau_2$, and $\tau_3$ for jets constructed from pileup per particle identification (PUPPI) particles (described in Sec.~\ref{sec:puppi}), using the $N$-subjettiness contrib package\,\cite{Thaler:2010tr,Thaler:2011gf} of \texttt{FastJet3}\,\cite{Cacciari:2011ma}. For the parameters, we use $\xi = 1.0$ with the unnormalised distance measure, i.e.~without the $1/d_0$ prefactor in Eq.~(\ref{eqn:nsubjettiness}). The subjet axes are found using the Winner-Take-All (WTA)\,\cite{Bertolini:2013iqa,Larkoski:2014uqa} $k_t$ axis-finding algorithm with one-pass minimisation. The WTA scheme provides an axis definition that is less sensitive to soft radiation, while the one-pass minimisation provides a computationally efficient procedure for optimising the axes\,\cite{Thaler:2011gf}. 

\subsection{Energy Correlation Functions}
\label{sec:ECF}

Following $N$-subjettiness, another class of JSS observables used to characterize the internal structure of jets is the ECFs\,\cite{Larkoski:2013eya,Larkoski:2014gra}. Energy correlation functions quantify the angular and momentum correlations among pairs or groups of jet constituents and are particularly effective for identifying jets originating from the hadronic decays of boosted heavy particles such as the Higgs boson or electroweak gauge bosons. Unlike $N$-subjettiness, which requires the identification of subjet axes, ECFs are constructed directly from the constituent information without relying on a specific axis definition.
This makes them useful for studying the internal structure of jets without introducing an additional axis-finding procedure.

The general form of an $N$-point ECF is given by
\begin{equation}
{\rm ECF}(N, \beta) = \sum_{i_1 < \cdots < i_N \in J} \left( \prod_{a=1}^{N} {p_T}_{i_a} \right) \left( \prod_{b=1}^{N-1} \prod_{c=b+1}^{N} \Delta R_{i_b i_c} \right)^{\beta},
\end{equation}
where the sum runs over all unique combinations of $N$ constituents in the jet. Here, ${p_T}_{i_a}$ is the transverse momentum of the ${i_a}^\text{th}$ constituent, and $R_{i_b i_c}$ denotes the angular distance in the $y$-$\phi$ plane between ${i_b}^\text{th}$ and ${i_c}^\text{th}$ constituents. The exponent $\beta$ controls the angular weighting. Larger values of $\beta$ emphasise wide-angle radiation, while smaller values increase the relative sensitivity to collinear configurations.
For $\beta > 0$, the ECFs are infrared and collinear (IRC) safe\,\cite{Larkoski:2013eya}.

The core idea of ECFs is to construct combinations of $p_T$-weighted angular distances between jet constituents. For example, the 1-point, 2-point, and 3-point ECFs, denoted by ECF$(1, \beta)$, ECF$(2, \beta)$, and ECF$(3, \beta)$, respectively, are given by
 
\begin{equation}
\label{eq:ecf}
\begin{aligned}
\mathrm{ECF}(1,\beta) &= \sum_{a\,\in J} \pta, \\
\mathrm{ECF}(2,\beta) &= \sum_{a<b\,\in J}
\pta\,\ptb \left(\dRab\right)^{\beta}, \\
\mathrm{ECF}(3,\beta) &= \sum_{a<b<c\,\in J}
\pta\,\ptb\,\ptc
\left(\Delta R_{ab}\,\Delta R_{ac}\,\Delta R_{bc}\right)^{\beta}.
\end{aligned}
\end{equation}
These functions are sensitive to the number and angular structure of hard prongs within the jet. 
In our study, we compute ECF$(1, \beta)$, ECF$(2, \beta)$, and ECF$(3, \beta)$ using the \texttt{FastJet3} contrib {\tt EnergyCorrelator}\,\cite{Cacciari:2011ma,Larkoski:2013eya,Larkoski:2014gra,Moult:2016cvt}.
Eventually, we construct the observables $C_1^{(\beta)}$, $C_2^{(\beta)}$\,\cite{Larkoski:2013eya}, and $D_2^{(\beta)}$\,\cite{Larkoski:2014gra}, which are used in the subsequent analysis. These observables are defined in terms of the ECFs as
\begin{equation}
   C_1^{(\beta)} = \frac{{\rm ECF}(2, \beta)}{[{\rm ECF}(1, \beta)]^2}, \ \ C_2^{(\beta)} = \frac{{\rm ECF}(3, \beta)\, {\rm ECF}(1, \beta)}{[{\rm ECF}(2, \beta)]^2}, \ \ D_2^{(\beta)} = \frac{{\rm ECF}(3, \beta)\,[{\rm ECF}(1, \beta)]^3}{[{\rm ECF}(2, \beta)]^3}.
   \label{eq:Ecorr}
\end{equation}
By construction, these observables are dimensionless and are also insensitive to recoil effects\,\cite{Larkoski:2013eya,Larkoski:2014gra,Moult:2016cvt}. The observable $C_1$, which contains a 2-point correlation function, is sensitive to the radiation pattern with respect to a single hard core in a jet and is particularly useful in the study of narrow jets, such as in quark versus gluon discrimination studies\,\cite{Larkoski:2013eya}. On the other hand, the $C_2$ and $D_2$ observables involve 3-point correlators and are useful for identifying jets with a 2-prong structure, such as jets originating from boosted gauge bosons or Higgs bosons, and for separating them from narrow QCD jets\,\cite{Larkoski:2014gra}.

\subsection{Jet Angularities}
\label{sec:JA}
Jet angularities are another class of JSS observables that characterise the internal angular structure of a jet.
They quantify the angular spread of the jet constituents around the jet axis and are therefore useful for distinguishing jets with different radiation patterns and characteristic angular sizes. In particular, they have been widely used to study the internal structure of quark- and gluon-initiated jets, as well as jets originating from boosted heavy particle decays.
Generalized jet angularities form a two-parameter family of observables, with one parameter controlling the weighting of the constituent transverse momentum and the other controlling the angular weighting\,\cite{Larkoski:2014pca,Gras:2017jty}.

For the present study, we consider the following form of the jet angularity:
\begin{equation}
e_\lambda = \sum_{a\,\in J} \frac{\pta}{\ptJ} \left(\dRJa\right)^\lambda,
\end{equation}
where $\pta$ is the transverse momentum of constituent $a$, $\ptJ$ is the transverse momentum of the jet, and $\dRJa$ is the angular distance between the constituent and the jet axis. The exponent $\lambda$ controls the angular weighting of the observable. The above definition corresponds to a particular choice within the generalised angularity family, where the transverse-momentum dependence is fixed to the linear $p_T$ fraction.

The generalised jet angularities defined in Refs.\,\cite{Larkoski:2014pca,Gras:2017jty} differ from the present definition also in the normalisation of the angular factor. In our case, we keep the angular factor unnormalised. This choice is because we plan to apply the angularity as a radius modifier of the dynamic radius algorithm. Since we want the modifier to retain information about the characteristic angular size of the jet, using the unnormalised angular measure allows the angular extent of the jet to be directly reflected in the resulting radius modifier.

\vspace{-4pt}
\section{JSS-Inspired Dynamic Radius Algorithm}
\vspace{-4pt}
\label{sec:drAlgo}
As discussed in Sec.~\ref{sec:jetAlgo}, traditional sequential jet clustering algorithms employ a fixed radius parameter, implying that all jets in an event are reconstructed with the same characteristic size. However, the physical size of jets can vary significantly depending on their origin and kinematics. Several approaches have therefore been proposed to incorporate this feature into jet reconstruction. For example, the Variable-$R$ jet algorithm\,\cite{Krohn:2009zg} allows the jet radius to vary during the clustering procedure, with the radius depending inversely on the transverse momentum of the jet. The variation is controlled by a maximum radius and a characteristic energy scale. The XCone algorithm\,\cite{Stewart:2015waa,Thaler:2015xaa} employs $N$-jettiness\,\cite{Stewart:2010tn} to determine the optimal jet axes to reconstruct a pre-defined number of jets from an event. 
Another approach, known as Fuzzy jets\,\cite{Mackey:2015hwa}, provides a probabilistic alternative to conventional jet clustering by assigning a probability for each particle to belong to multiple jets. Other related approaches include SHAPER\,\cite{Ba:2023hix} and PAIReD jets\,\cite{Mondal:2023law}.

Our previous proposal, the dynamic radius (DR) jet algorithm\,\cite{Mukhopadhyaya:2023rsb}, provides another approach in which the effective radius of a jet is determined dynamically during the clustering procedure. In Ref.\,\cite{Mukhopadhyaya:2023rsb}, the idea of dynamic radius jet clustering was introduced, where the radius associated with each jet is allowed to evolve during the clustering sequence. Starting from an initial radius parameter $R_0$, the radius of an evolving proto-jet is modified iteratively according to its internal substructure. In this approach, the dynamic radius of the $i^{\rm th}$ jet/proto-jet is expressed as
\begin{equation}
R_d = R_0 + \sigma_i, \label{eqn:rd}
\end{equation}
where $\sigma_i$ denotes a radius modifier determined from the constituents already clustered into the proto-jet. In that implementation, the distance measures considered are as given in Eqs.~(\ref{eqn:dij})--(\ref{eqn:diB}) by replacing $R\to \Rd$, and the modifier was defined through the variance of the pairwise angular separation among constituents,
\begin{equation}
\label{eqn:var}
\sigma_i^2 = \dfrac{\sumab \pta\,\ptb\, \dRab^2}
{\sumab \pta\,\ptb} -
\left(
\dfrac{\sumab \pta\,\ptb\, \dRab}
{\sumab \pta\,\ptb}
\right)^2,
\end{equation}
with $a$ and $b$ running over all constituents of the proto-jet. Consequently, the jet radius is no longer universal but is determined dynamically according to the structure of the jet being formed.

Interestingly, the above quantity can be related to the two-point energy correlation function 
observable $C_1^{(\beta)}$\,\cite{Larkoski:2013eya} through $\sigma_i^2 = C_1^{(2)} - \left(C_1^{(1)}\right)^2$.
This relation suggests that the radius modifier is intrinsically connected to the substructure properties of the jet. For a jet consisting of two hard prongs separated by an angle $\theta$, the energy correlators scale approximately as $C_1^{(\beta)} \sim z\,\theta^\beta$,
where $z$ denotes the energy fraction carried by the softer prong\,\cite{Larkoski:2013eya}. Consequently, one finds $\sigma_i \propto \theta$, indicating that the radius modifier is directly sensitive to the characteristic opening angle of the underlying splitting or decay\,\cite{Ghosh:2025gdq}. Jets originating from boosted heavy particles, such as electroweak gauge bosons or top quarks, generally exhibit larger angular separations among their decay products and therefore tend to acquire larger effective radii.
On the other hand, QCD jets initiated by light quarks or gluons are typically dominated by collinear radiation. In this case, the two-point correlator scales as $C_1^{(2)}\sim m^2/p_T^2$, leading to smaller values of the radius modifier\,\cite{Proceedings:2018jsb}. As a result, narrow QCD jets naturally favour smaller effective radii. The original dynamic radius clustering prescription, therefore, provides a mechanism to adapt the jet size according to the characteristic angular size associated with the jet.

The above observations also indicate that certain JSS observables encode valuable information about the angular extent of a jet. Since observables such as energy correlation functions\,\cite{Larkoski:2013eya} and jet angularities\,\cite{Berger:2003iw,Almeida:2008yp,Larkoski:2014pca,Andersen:2016qtm} are sensitive to the radiation pattern and prong structure of a jet, they can help distinguish fat jets arising from boosted heavy-particle decays from narrow jets initiated by light quarks and gluons. Motivated by this connection between JSS and jet formation, in the present work we generalise our previous study\,\cite{Mukhopadhyaya:2023rsb} of the dynamic radius clustering framework by going beyond a single choice of radius modifier. In the previous work, only one specific radius modifier was considered, as given by Eqs.~(\ref{eqn:diB}) and (\ref{eqn:dijtilde}). In the present work, we construct a variety of radius modifiers based on observables sensitive to the angular structure and prong structure of the jet with the aim of assigning larger radii to fat jets while retaining smaller radii for narrow jets, thereby providing a more adaptive reconstruction of jets with different characteristic angular scales within the same event.

As discussed in Sec.~\ref{sec:jetAlgo}, the modification to the radius parameter, $R\to \Rd$, to incorporate dynamic radius can be implemented either in the $\diB$ definition [Eq.~(\ref{eqn:diB})] or in the
$\dij$ definition [Eq.~(\ref{eqn:dijtilde})]. For easy reference, we call an algorithm `Type I' if the
dynamic radius dependence is incorporated in the $\diB$ definition. Otherwise, it is referred to as a
`Type II' algorithm. 
A list of these different 
modifications is given in Table~\ref{tab:dyn-seq-algo}. We further note that the distance measures retain the same form as those used in the $k_t$-type sequential recombination algorithms. Following the generalised nomenclature, we therefore denote the corresponding dynamic radius algorithms as DR-AK, DR-CA, and DR-KT, corresponding to the exponent $n=-1$, 0, and 1, respectively. 

\begin{table}[!h]
\renewcommand{\arraystretch}{2.2}
\begin{center}
\begin{tabular}{|C{0.42\textwidth}|c|c|c|}
\hline
Algorithm & Type & $\dij$ & $\diB$ \\ 
\hline
\hline
\multirow{2}{0.40\textwidth}{Dynamic radius $\kt$-type algorithms. The acronyms
are DR-AK, DR-CA, and DR-KT for $n=-1$, 0, and 1, respectively.} & I &
$\min\left(\pti^{2n},\ptj^{2n}\right) \dRij^2$ & $\pti^{2n} \Rd^2$ \\
\cline{2-4}
& II & $\min\left(\pti^{2n},\ptj^{2n}\right) \left(\dfrac{\dRij}{\Rd}\right)^2$
& $\pti^{2n}$ \\[4pt]
\hline
\end{tabular}
\caption{Definitions of the pairwise ($\dij$) and beam ($\diB$) distance measures for Type I and Type II dynamic-radius algorithms.
The different choices for $\Rd$ are listed in Tables~\ref{tab:modifier} and \ref{tab:Rd-choices}.}
\vspace{-16pt}
\label{tab:dyn-seq-algo}
\end{center}
\end{table}

In the proposed formalism, we consider different ways of adding a dynamic growth
rate to the starting radius $\Rnot$. For example, the addition of a dynamic part to
$\Rnot$ can be done either in a linear fashion or in quadrature. Since this is an
addition to $\Rnot$, the dimension of the parameter should also be the same as the
dimension of the radius. Furthermore, the evolution of the radius is expected to be governed by the
internal structure of each evolving proto-jet. Therefore, two commonly used statistical measures, {\it viz.} average ($\Ravgi$) and standard deviation ($\sigmaRi$) of
$\dR$, can be used to quantify the growth of each proto-jet.
These averages or standard deviations can be constructed either from the pairwise angular separation between jet constituents or from the angular distance between the jet axis and its constituents. If the weights of the average or standard deviations are $p_T$ of the jet constituents, the corresponding quantities are directly related to the ECFs\,\cite{Larkoski:2013eya} in the former case, whereas in the latter case they are closely connected to JAs\,\cite{Berger:2003iw,Almeida:2008yp,Larkoski:2014pca,Andersen:2016qtm}. The most general expressions for these two classes of quantities are listed in Table~\ref{tab:modifier}.

\begin{table}[!h]
\begin{center}
\begin{tabular}{|c|c|c|}
\hline
Radius & \multirow{2}{*}{ECF} & \multirow{2}{*}{JA} \\
~modifier~ & & \\
\hline
 & & \\[-12pt]
$\Ravgi$     & $\langle\dRab \rangle = C_1^{(1)}$ & $\langle\dRJa\rangle = e_1$ \\[6pt]
\hline
 & & \\[-12pt]
$\sigmaRi^2$ & $\langle\dRab^2 \rangle - \langle\dRab \rangle^2 = C_1^{(2)} - \left(C_1^{(1)}\right)^2$ & $\langle\dRJa^2\rangle - \langle\dRJa \rangle^2 = e_2 - \left({e_1}\right)^2$ \\[8pt]
\hline
\end{tabular}
\caption{Radius modifiers based on the mean and variance of constituent-based and jet-axis-based observables. Here, $a,b$ label the jet constituents and $J$ denotes the jet axis. The terms ECF and JA denote energy correlation functions and jet angularities, respectively, with $C_1^{(1,2)}$ defined in Sec.~\ref{sec:ECF} and $e_{1,2}$ defined in Sec.~\ref{sec:JA}.} 
\label{tab:modifier}
\end{center}
\end{table}

Furthermore, a constant or proto-jet dependent factor 
($\alpha_i$) may also be used to appropriately scale the additive part
($\Ravgi$ or $\sigmaRi$) in $\Rdi$ for $\ith$ proto-jet.
In a nutshell, in this general formalism, the average ($\Ravgi$) or the
standard deviation ($\sigmaRi$) of $\dR$ between the constituents or between the jet axis
and constituents can be added to $\Rnot$ linearly or in quadrature along with a
scale factor. An extensive list of the most general variations is listed in
Table~\ref{tab:Rd-choices}.
\begin{table}[!h]
\renewcommand{\arraystretch}{1.5}
\begin{center}
\begin{tabular}{|c|c|c|}
\hline
~~Variations~~ & $\Rd^2\ \text{in}\ \dij$ & $\Rd^2\ \text{in}\ \diB$ \\ 
\hline
\hline
A & ~~$\left(\Rnot + \alpha_i \sigmaRi + \alpha_j \sigmaRj \right)^2$~~ & ~~$\left(\Rnot + \alpha_i \sigmaRi \right)^2$~~ \\
\hline
B & $\left(\Rnot^2 + \alpha_i^2 \sigmaRi^2 + \alpha_j^2 \sigmaRj^2 \right)$ & $\left(\Rnot^2 + \alpha_i^2 \sigmaRi^2 \right)$ \\
\hline
C & $\left(\Rnot + \alpha_i\,\Ravgi + \alpha_j\,\Ravgj \right)^2$ & $\left(\Rnot + \alpha_i\,\Ravgi\right)^2$ \\
\hline
D & $\left(\Rnot^2 + \alpha_i^2\,\Ravgi^2 + \alpha_j^2\,\Ravgj^2\right)$ & $\left(\Rnot^2 + \alpha_i^2\,\Ravgi^2\right)$ \\
\hline
\hline
\end{tabular}
\caption{A comprehensive list of variations by which the dynamicity of the radius parameter of an evolving proto-jet can be added.}
\vspace{-20pt}
\label{tab:Rd-choices}
\end{center}
\end{table}

In our previous study\,\cite{Mukhopadhyaya:2023rsb}, the modification was performed using a specific method {\it viz.} the linear addition of $\sigma_i$ with coefficient $\alpha_i=1$ in the Type-I scenario. Here, however, we plan to examine it in a more general fashion, {\it i.e.}, proposing some alternate variations around the same concept. A comprehensive study of these proposals is the primary motivation of this paper. 
We further note that the radius modifiers are related to IRC-safe JSS observables ECFs or JAs. 
Therefore, the introduction of these radius modifiers does not lead to any additional IRC safety issues. Additionally, the form of $\dij$ and $\diB$ ensures that the IRC safety properties are inherited from the underlying fixed radius clustering algorithms.

In the following, we adopt the
{\tt <Type><Variation>-<radius modifier>} nomenclature
to denote a particular variant of the dynamic radius algorithms. For example, the algorithm corresponding to Type I, variation A, and a radius modifier constructed from ECFs will be denoted by {\tt IA-ECF}.

\subsection{Implementation within FastJet3 Framework}
We have implemented different variants, as discussed in Sec.~\ref{sec:drAlgo}, of the dynamic radius jet clustering algorithm as a \texttt{FastJet3} plugin\,\cite{Cacciari:2011ma}. \texttt{FastJet3} framework provides a flexible framework for implementing jet algorithms, including classes for handling distance measures and clustering sequences. Our implementation makes use of the \texttt{NNBase} and \texttt{NNH} classes to handle pairwise and beam distance computations. Our $\dij$ measures are symmetric in the $i$ and $j$ indices as required by these classes. 

The clustering procedure is performed using the built-in \texttt{ClusterSequence} class of \texttt{FastJet3}, while the kinematic properties of particles and intermediate proto-jets are handled using the \texttt{PseudoJet} class. We make use of the \texttt{user\_info} feature of \texttt{PseudoJet} to attach a custom \texttt{DynamicRJetInfo} structure that stores pseudojet properties such as the mean angular spread and the mean square radius. These quantities are used to compute a jet-specific dynamic radius modifier, which subsequently enters into $\dij$ and $\diB$ calculations depending on the variant used (Type I or II, and A, B, C or D). The effective radius of each evolving jet is evaluated from its internal substructure during the clustering sequence. The final value of the dynamic radius is stored and can therefore be directly accessed for further analysis.

As discussed previously, the parameter $\alpha_i$ controls the growth of the dynamic radius and acts as a scaling factor for the internal angular spread of $\ith$ proto-jet. By adjusting $\alpha_i$, one can regulate how sensitively the jet radius responds to the substructure of the evolving jet. In our implementation, we examine multiple choices of $\alpha_i$ to study their impact on jet reconstruction and discrimination power. These include fixed values such as $\alpha_i = 1$ and $\alpha_i = 2$, as well as a dynamic, kinematics-dependent option defined as $\alpha_i = \dfrac{2 m_i}{\pti \Rnot}$, where $m_i$ and $\pti$ are the mass and transverse momentum of the evolving pseudojet. The motivation behind the dynamic choice is to allow the radius growth to be sensitive not only to geometric spread but also to the jet energy scale and invariant mass.

\section{Analysis procedure}
\label{sec:workflow}
\subsection{Event Generation}
We study the signal processes $ pp \to Vj $ ($V=W^\pm/Z$) and $ pp \to tj $ to test the efficacy of the dynamic radius jet clustering algorithm in events containing a boosted fat jet produced in association with a narrow jet. To ensure boosted kinematics, two benchmark samples are generated with $Vj$ final states with $p_T^V > 300~\mathrm{GeV}$ and $p_T^V > 500~\mathrm{GeV}$, while one benchmark sample is generated for the $tj$ signal with $p_T^t > 500~\mathrm{GeV}$. As background, we consider QCD dijet production with the same $p_T$ thresholds, {\it i.e.}, with $p_T^j > 300~\mathrm{GeV}$ and $p_T^j > 500~\mathrm{GeV}$. The description of these samples, along with the labels to be used further, is given in Table~\ref{tab:samples}.
\begin{table}[!h]
\begin{tabular}{l p{4.5in}}
\hline\hline
\quad \bf Sample\quad & \centerline{\bf Description} \\ [-16pt]
\hline
\quad \Vjth \qquad\qquad & This sample probes the moderately boosted region of the $Vj$ process, with the vector boson satisfying $p_T^V>300~\mathrm{GeV}$. The vector boson is allowed to decay only through hadronic modes. \\
\quad \Vjfh \qquad\qquad & This sample corresponds to the highly boosted region of the $Vj$ process, with $p_T^V>500~\mathrm{GeV}$. \\
\quad \tjfh \qquad\qquad & This sample corresponds to the boosted region of the $tj$ process, with the top quark satisfying $p_T^t>500~\mathrm{GeV}$.\\
\hline
\quad \jjth\qquad\qquad & This sample is QCD dijet events generated with $p_T^j>300~\mathrm{GeV}$.\\
\quad\jjfh\qquad\qquad & This sample is QCD dijet events generated with $p_T^j>500~\mathrm{GeV}$.\\
\hline\hline
\end{tabular}
\caption{Signal and background samples and their descriptions.}
\label{tab:samples}
\end{table}

We generate $10^5$ signal events and $10^6$ background events for each benchmark point. Event generation is performed using \texttt{MadGraph5\_aMC@NLO}\,\cite{Alwall:2014hca} at $\sqrt{s}=14~\mathrm{TeV}$ LHC at the parton-level, followed by showering and hadronisation in \texttt{PYTHIA 8.3}\,\cite{Sjostrand:2006za,Sjostrand:2007gs,Sjostrand:2014zea,Bierlich:2022pfr}. Pileup interactions with mean $n_{\text{PU}} = 150$ are then merged with the samples along with the detector simulation using \texttt{Delphes 3.5.0}\,\cite{deFavereau:2013fsa}. To mitigate pileup, we use the PUPPI algorithm\,\cite{Bertolini:2014bba} (discussed briefly in Sec.~\ref{sec:puppi}). Jets are reconstructed with \texttt{FastJet3}\,\cite{Cacciari:2011ma} using both standard and dynamic radius algorithms with the PUPPI particles.

For the purpose of object identification and truth labelling, a geometrical matching between reconstructed jets and parton-level objects is performed. After clustering, only the two leading jets with $p_T > 150~\mathrm{GeV}$ are considered. For each jet, the angular distance $\Delta R$
is computed with respect to both the parton-level gauge boson ($V = W^\pm/Z$) or top quark, and the associated parton. For truth labelling, a matching criterion of $\Delta R < 0.5$ between parton-level objects and reconstructed jets is imposed to establish the correspondence. Events that fail to satisfy the matching criteria are discarded to ensure a consistent and unambiguous association between reconstructed jets and their underlying parton-level origins.

\begin{figure}[!h]
    \begin{center}
    \includegraphics[clip, trim=3cm 15cm 3cm 2cm, width=0.9\textwidth]{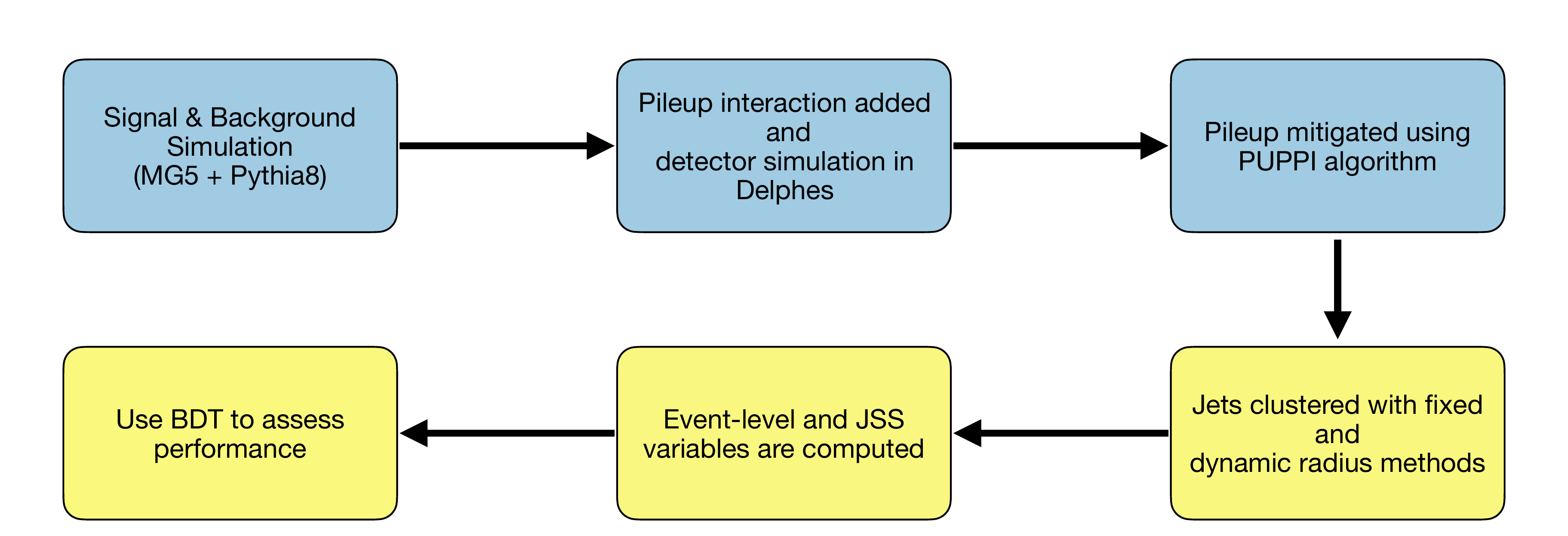}
    \end{center}
    \caption{Schematic diagram representing the workflow followed for the analysis. }
    \label{fig:Workflow}
\end{figure}

After the jet identification, Soft Drop grooming\,\cite{Larkoski:2014wba} is performed with an energy fraction regulator $z_\text{cut} = 0.1$ and angular exponent $\beta_\text{SD} = 1.0$. The JSS observables $\tau_{21}, \tau_{32}, e_2, e_3, C_2, D_2$ are then constructed from the Soft Drop jets. In addition, the jet mass, energy, transverse momentum $p_T$, and pseudorapidity $\eta$ are computed from the Soft Drop jet. The final jet radius, however, is taken from the original jet before grooming. These observables and event-level variables are used as inputs to a Boosted Decision Tree (BDT) classifier to discriminate signal from background. The overall workflow is illustrated in Figure~\ref{fig:Workflow}. 

\subsection{PileUp Per Particle Identification}
\label{sec:puppi}
In hadron collider environments like the one seen at the LHC, multiple proton-proton interactions often occur in the same bunch crossing, leading to what is known as \emph{pileup}. These additional soft interactions can significantly contaminate the event by introducing extra particles that are unrelated to the hard scattering process of interest. This contamination affects the reconstruction of key observables such as jet momentum. The PileUp Per Particle Identification (PUPPI) algorithm was proposed to mitigate these effects by evaluating each particle individually and determining its likelihood of originating from the primary interaction rather than from pileup\,\cite{Bertolini:2014bba}.

The PUPPI algorithm calculates a local shape variable, $\alpha$, for each particle to characterise its surrounding radiation pattern based on nearby particles' angular and momentum features. Particles from the primary vertex typically have collimated, energetic neighbours, while pileup particles appear more diffuse. By comparing $\alpha$ to the distribution for known pileup particles, the algorithm assigns each particle a weight between 0 and 1, indicating its likelihood of originating from the primary interaction. Once the PUPPI weights are calculated, they are used to rescale the four-momentum of each particle. Particles likely from pileup (weights close to 0) are suppressed or removed, while those from the primary interaction (weights close to 1) are retained with minimal modification. This particle-level cleaning results in significantly improved performance for jet reconstruction and JSS analysis. 

In our methodology, we utilise the PUPPI module configured with parameters aligned closely with the CMS default settings\,\cite{CMS-PAS-JME-14-001}. This implementation is done using a \texttt{Delphes} card\,\cite{deFavereau:2013fsa}. A cone of radius $R = 0.2$ is used to compute local shape variables, and a minimum transverse momentum threshold of $p_T > 0.1$ GeV is applied to suppress soft noise. The algorithm is run with Charged Hadron Subtraction (CHS) enabled and tuned separately for two pseudorapidity regions: $0.0 < |\eta| < 4.0$ (central) and $4.0 < |\eta| < 5.0$ (forward), to account for detector granularity. We also apply a minimum PUPPI weight threshold of 0.05, below which particles are discarded to minimise residual contamination from pileup. The PUPPI-corrected particles are then used in jet clustering and substructure analysis.

\subsection{BDT}
\label{sec:bdt}
To assess the performance of the variants of dynamic radius jet clustering algorithms introduced in the previous section, we adopt a supervised learning approach using a Boosted Decision Tree (BDT) classifier~\cite{63451,FREUND1995256,Friedman:2001wbq,TMVA:2007ngy}. The BDTs are widely used in particle physics analyses for their ability to capture non-linear correlations among observables. In our context, the BDT is trained to discriminate jets originating from hadronic decays of boosted vector bosons/ top quarks (signal) from QCD jets (background), using a set of jet and JSS observables. These input features used in our analysis are tabulated in Table~\ref{tab:bdt_inputs} for both the analyses -- (i) $\Vj$ signal when the signal contains a boosted $W^\pm$ or $Z$, and (ii) $\tj$ analysis when the signal contains a boosted top quark. For each event, observables are extracted both from the primary (fat) jet and its associated jet.

\begin{table}[htbp]
\centering
\newcolumntype{C}[1]{>{\centering\arraybackslash}m{#1}}
\begin{tabular}{|C{7.9cm}|C{2.5cm}|C{2.5cm}|}
\hline
\textbf{Variables} & \textbf{\Vj signal} & \textbf{\tj signal} \\
\hline
Soft Drop jet mass ($m_J$) & \checkmark & \checkmark \\
Soft Drop jet energy ($E_J$) & \checkmark & \checkmark \\
Soft Drop jet eta ($\eta_J$) & \checkmark & \checkmark \\
Soft Drop jet $p_T$ & \checkmark & \checkmark \\
Final dynamic jet-radius ($R_d$) & \checkmark & \checkmark \\[4pt]
$\sum k_t$ & \checkmark & \checkmark \\
$\tau_{21}$ & \checkmark & -- \\
$\tau_{32}$ & -- & \checkmark \\
$C_2^{(\beta)}$ [$\beta=1.0$] & \checkmark & -- \\
$D_2^{(\beta)}$ [$\beta=1.0$] & \checkmark & -- \\
\hline
\end{tabular}
\caption{Input variables used for BDT training. A check mark indicates that the observable is included in the respective BDT. The variable $R_d$ is used only for dynamic radius jets. The variable $\sum k_t$ denotes the sum of the $k_t$ distances of the jet constituents from the jet axis. The parameter choices for $\tau_{21}$ and $\tau_{32}$ are given in Sec.~\ref{sec:nsubjettiness}.}
\label{tab:bdt_inputs}
\end{table}

The BDT is implemented using the \texttt{XGBoost} library\,\cite{Chen_2016}. The trained BDT model outputs probability scores for each jet being signal-like or background-like. These scores are then used to construct Receiver Operating Characteristic (ROC) curves.
This framework allows us to compare how different dynamic radius clustering variants enhance signal-to-background separation and whether our proposed variants of clustering improve tagging efficiency over the conventional anti-$k_t$ algorithm.

\section{Results and Discussion}
\label{sec:result}

In this section, we present the results obtained using different variants of the dynamic radius jet clustering algorithm discussed in Sec.~\ref{sec:drAlgo}. The objective is to study the performance of these variants in distinguishing signal events containing a fat jet originating from a boosted heavy particle in association with a narrow jet from the  QCD dijet background, which predominantly consists of narrow jets. To illustrate the performance, we consider the signal processes: $\bullet$ $pp \rightarrow Vj$ ($V = W$ or $Z$), and $\bullet$~$pp \rightarrow tj$. 

All simulated samples are processed through the workflow described in Sec.~\ref{sec:workflow}. For the comparative study, we consider $\jj$ production as the background process. Since the analysis is performed in three different kinematic regions, namely $\Vjth$, $\Vjfh$, and $\tjfh$, different background samples are used accordingly. The $\jjth$ sample is used as the background for the $\Vjth$ analysis, while the $\jjfh$ sample is used as the background for both the $\Vjfh$ and $\tjfh$ analyses. In the following, we present the results for these three signal samples separately.

We further note that we have tested all the variants listed in Sec.~\ref{sec:drAlgo}. For brevity, we present a subset of four variants that show the best performance. The remaining variants are also implemented within our framework and will be available in the released version of the code, following the convention and nomenclature defined in Sec.~\ref{sec:drAlgo}.

\subsection{Analysis of $\Vjth$ vs.~\jjth Samples}
\label{sec:Vj300-jj300}
\begin{figure}[!b]
\begin{center}
\includegraphics[width=0.48\textwidth]{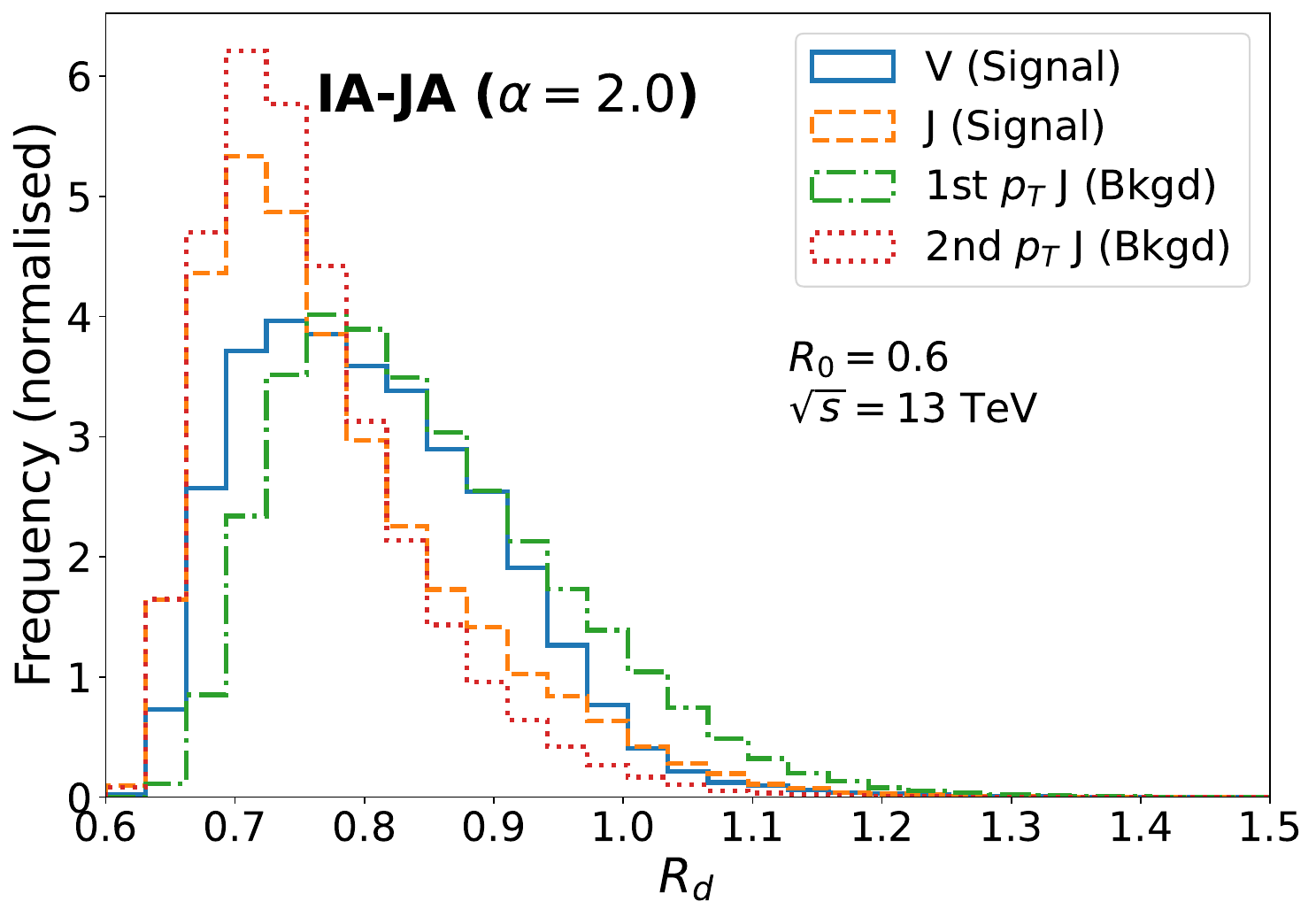}
\includegraphics[width=0.48\textwidth]{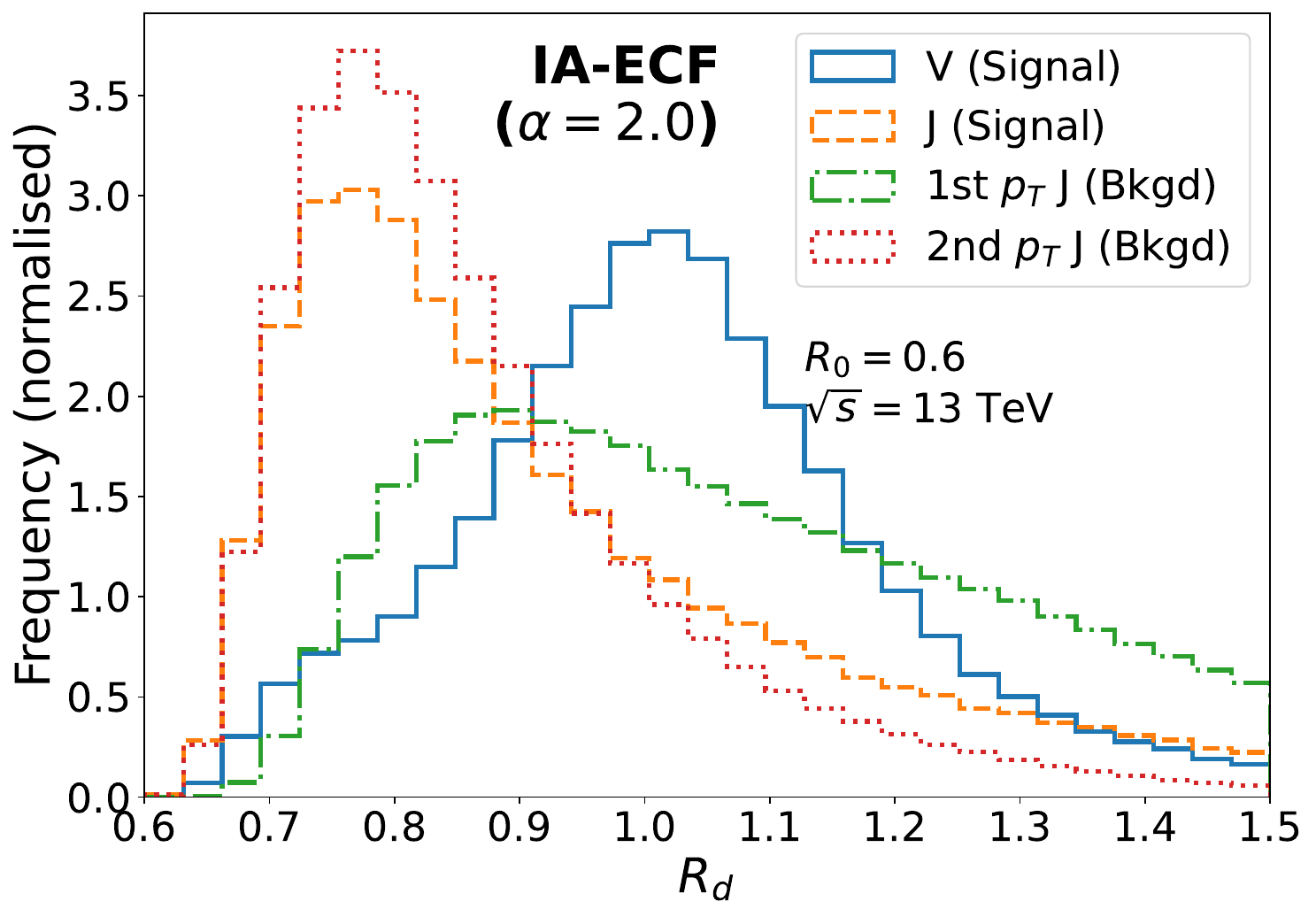}
\includegraphics[width=0.48\textwidth]{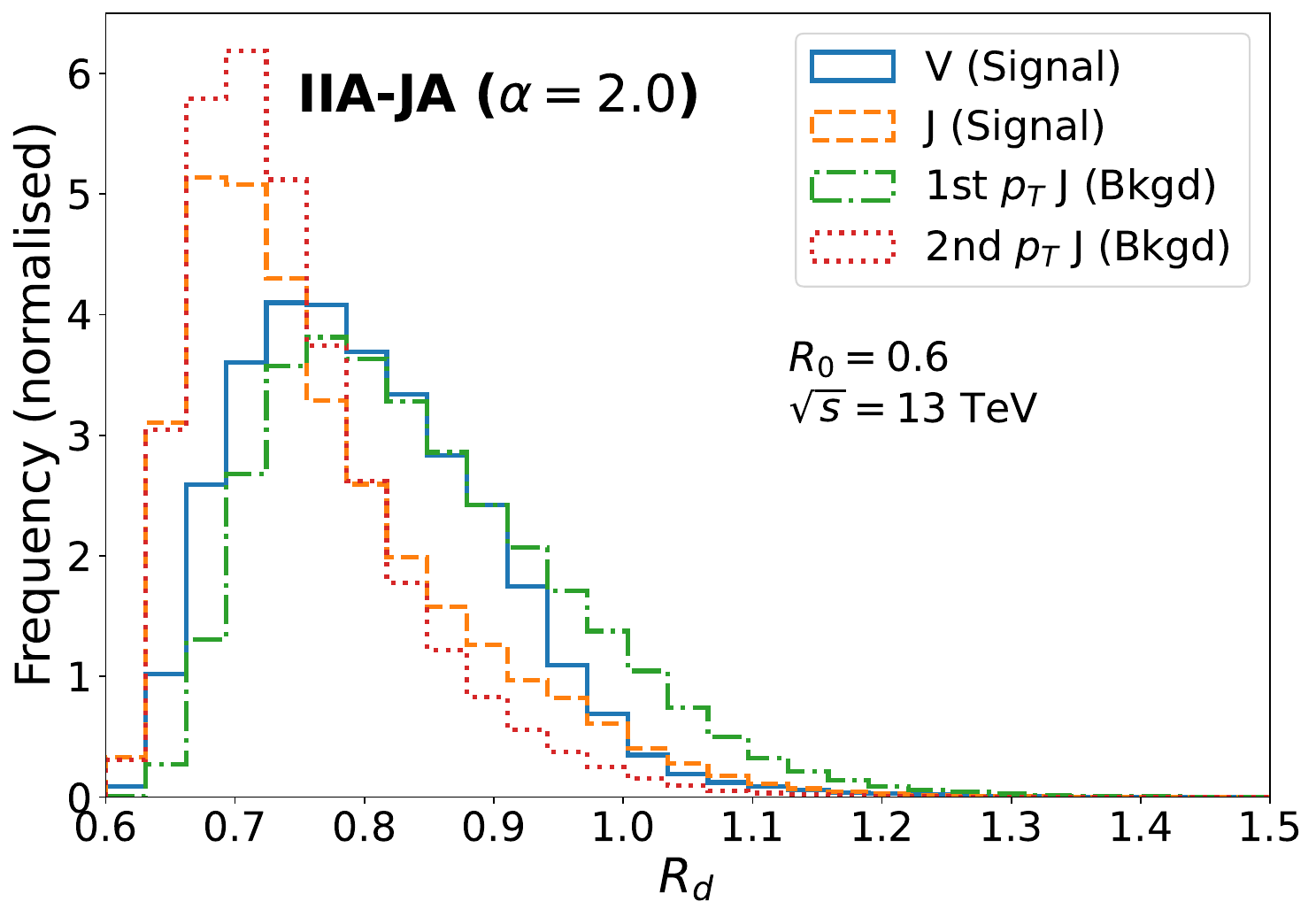}
\includegraphics[width=0.48\textwidth]{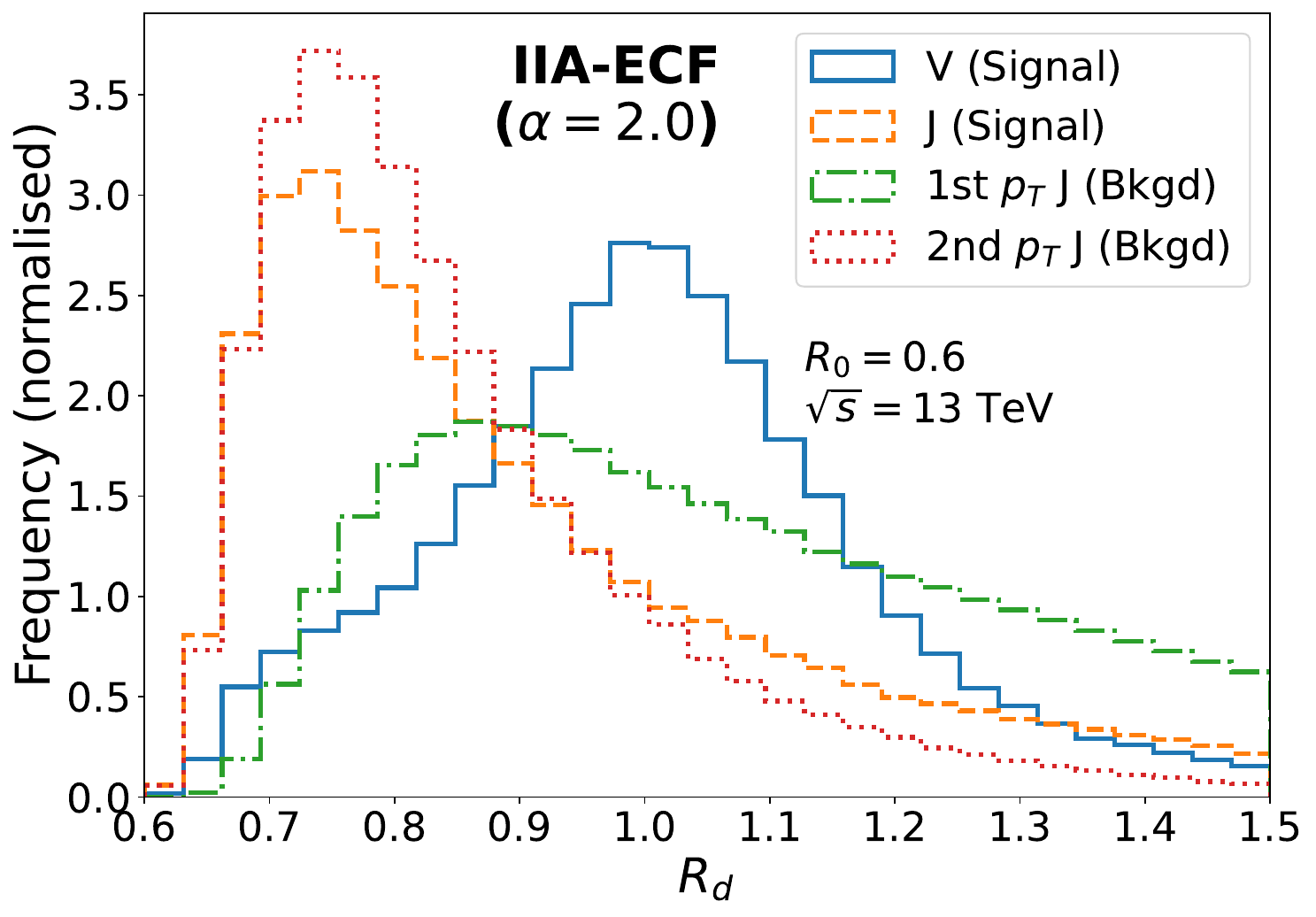}
\end{center}
\vspace{-20pt}
\caption{Normalised distributions of $R_d$ for different DR variants for the $\Vjth$ (signal) and $\jjth$ (background) samples with starting radius $R_0=0.6$.
Each panel shows a different DR variant, as indicated in the panel, with the variants described in Sec.~\ref{sec:drAlgo}. ECF and JA denote energy correlation functions and jet angularities, respectively.}
\label{fig:Rd-Vj300-jj}
\end{figure}

We begin with the $\Vjth$ sample, where the vector boson satisfies $p_T^V>300~\mathrm{GeV}$ at the parton-level. Figure~\ref{fig:Rd-Vj300-jj} shows the distributions of the final dynamically adjusted jet radius for the identified jets originating from the vector boson and the accompanying light-flavour jet. Results are presented for the four dynamic radius prescriptions, namely {\tt IA-JA}, {\tt IA-ECF}, {\tt IIA-JA}, and {\tt IIA-ECF}.
For comparison, the corresponding radius distributions for the leading and subleading jets in the $\jjth$ background sample are also shown. The initial radius parameter is chosen to be $R_0=0.6$. The Soft Drop jet mass distributions are provided in Appendix~\ref{sec:jetmass}.

As expected, jets originating from boosted vector-boson decays are assigned larger radii, reflecting their wider two-prong substructure. In contrast, the accompanying light jet in the signal events and the jets in the background $\jjth$ sample tend to acquire smaller effective radii. This demonstrates that the dynamic radius algorithm adapts the jet radius according to the underlying jet structure, allowing jets with different characteristic angular scales to be reconstructed with different effective radii.

\begin{figure}[!b]
\begin{center}
\includegraphics[width=0.48\textwidth]{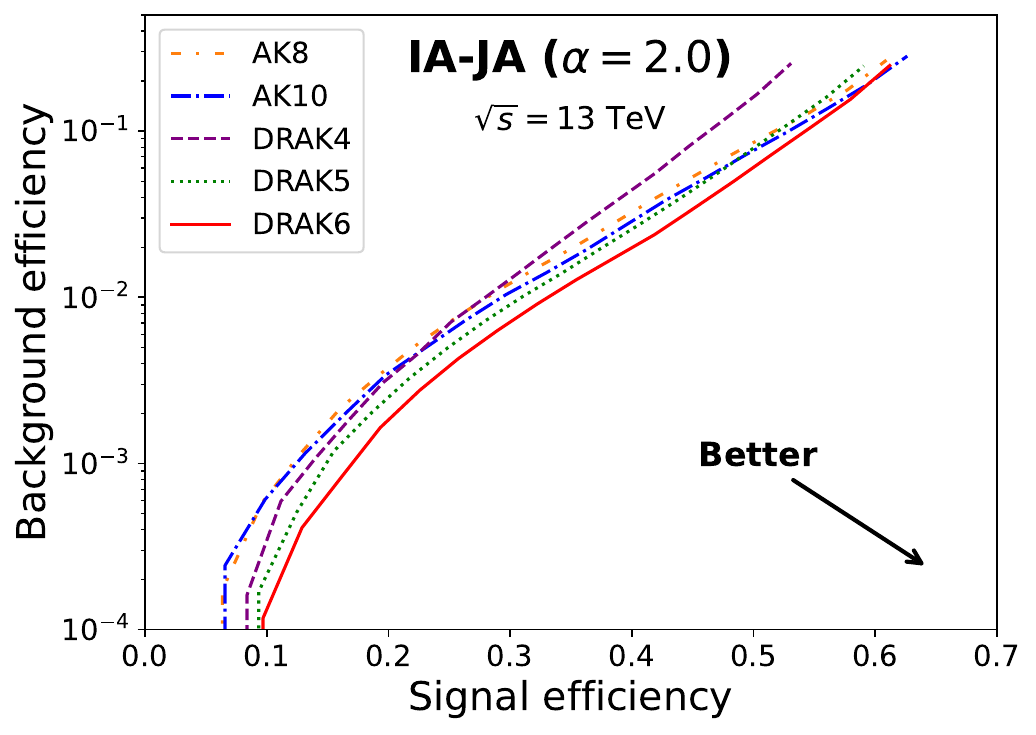}\hfill
\includegraphics[width=0.48\textwidth]{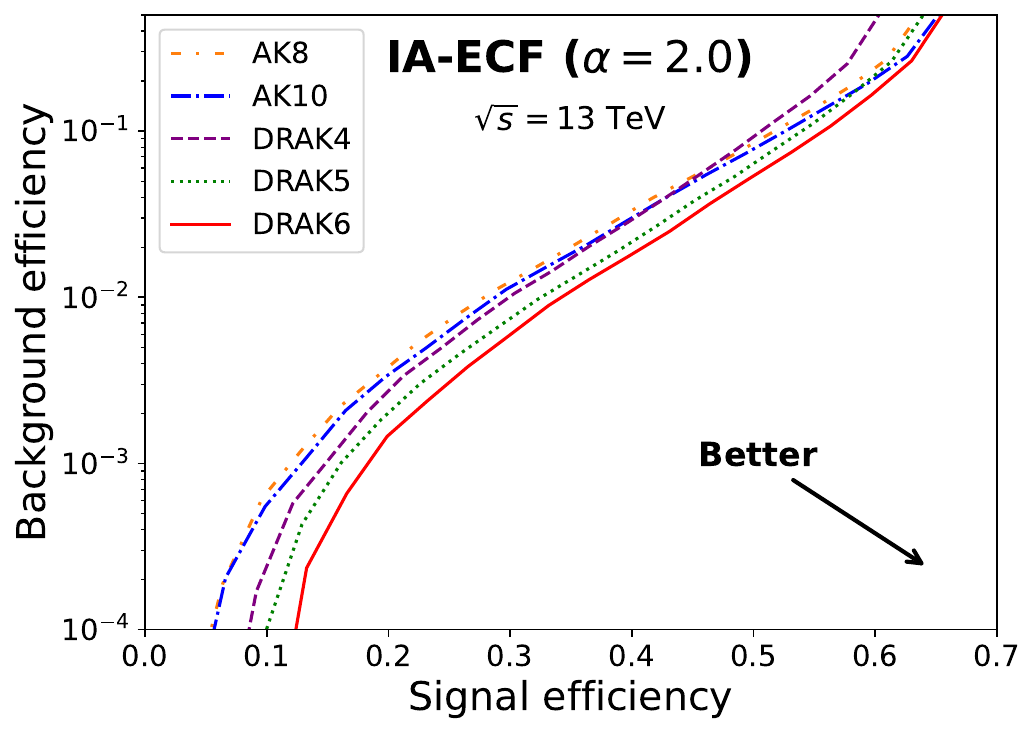}
\includegraphics[width=0.48\textwidth]{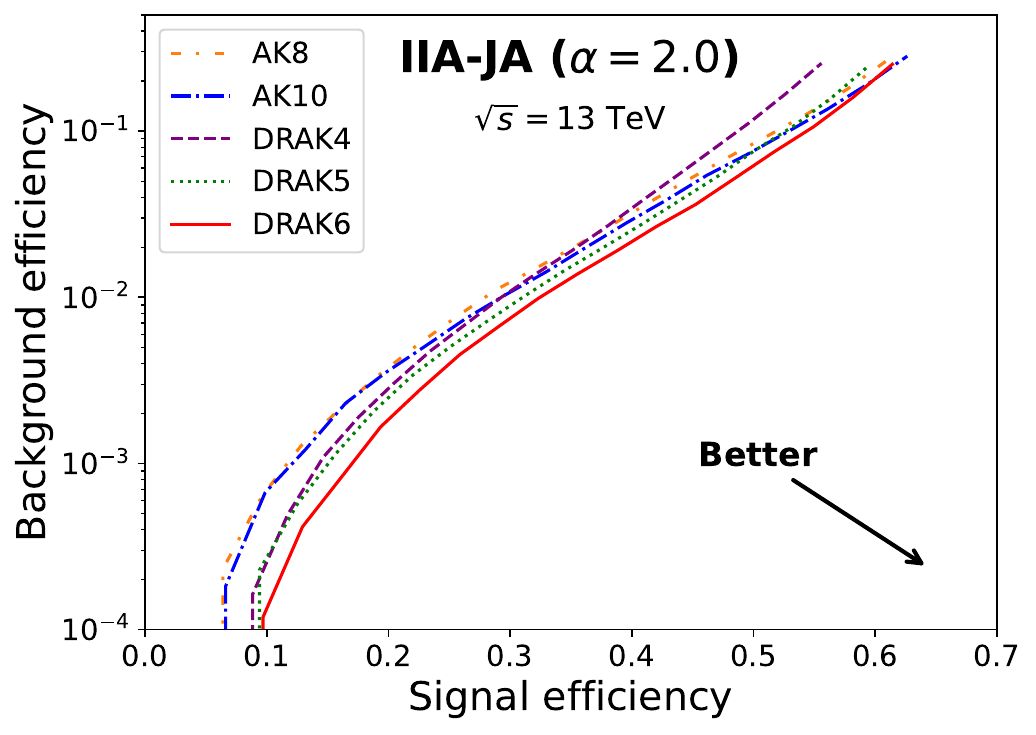}\hfill
\includegraphics[width=0.48\textwidth]{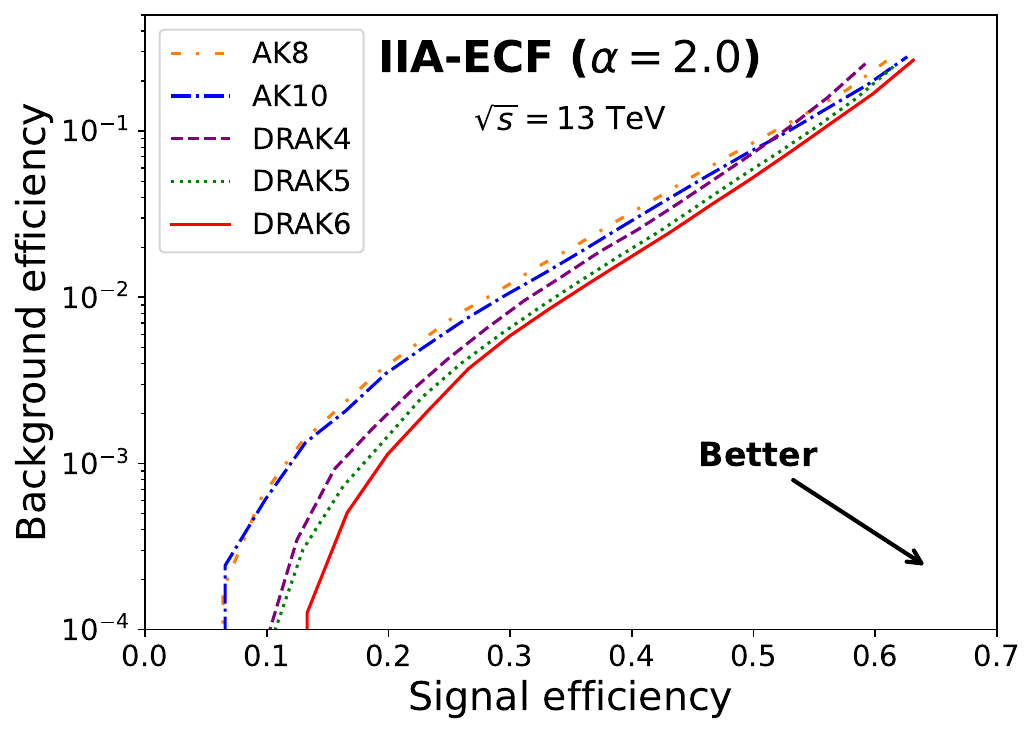}
\end{center}
\vspace{-20pt}
\caption{Comparison of ROC curves for different clustering methods for \Vjth (signal) vs.~\jjth (background) samples. AK8 and AK10 correspond to fixed radius anti-$k_t$ algorithms with $R=0.8$ and 1.0, respectively, while DRAK4, DRAK5, and DRAK6 correspond to dynamic radius anti-$k_t$ algorithms with starting radius $R_0=0.4$, 0.5, and 0.6, respectively. Each panel corresponds to one DR variant, as indicated in the panel, with the variants described in Sec.~\ref{sec:drAlgo}. ECF and JA denote energy correlation functions and jet angularities, respectively.}
\label{fig:Vj300-jj}
\end{figure}

To quantify the impact of dynamic radius reconstruction on signal-background discrimination, we perform a multivariate analysis using the observables described in Sec.~\ref{sec:bdt}. The resulting ROC curves are shown in Figure~\ref{fig:Vj300-jj}. Dynamic radius jets with $R_0=0.6$ are studied for three different initial radius parameters, $R_0=0.4$ (DRAK4), $0.5$ (DRAK5), and $0.6$ (DRAK6). Among these choices, the configuration with $R_0=0.6$ provides the best overall performance. For comparison, the same analysis is also performed using conventional anti-$k_t$ jets with fixed radius parameters $R=0.8$ (AK8) and $R=1.0$ (AK10). The corresponding ROC curves are shown in the same figure. We observe that the dynamic radius jets consistently provide better signal-background discrimination than the fixed radius anti-$k_t$ jets, resulting in improved signal efficiency for a given background rejection rate.

A comparison of the four variants shown in Figure~\ref{fig:Vj300-jj} allows us to assess the impact of the radius modifier and the dynamic-radius construction. Comparing the left and right columns, the variants based on the ECF-inspired radius modifier show better performance than those based on the JA-inspired modifier. Similarly, comparing the top and bottom rows indicates that the type IIA construction performs better than type IA. Overall, the IIA-ECF variant provides the best performance among the four variants considered.

\vspace{-4pt}
\subsection{Analysis of $\Vjfh$ vs.~$\jjfh$ Samples}
\label{sec:Vj500-jj500}

\begin{figure}[!b]
\begin{center}
\includegraphics[width=0.48\textwidth]{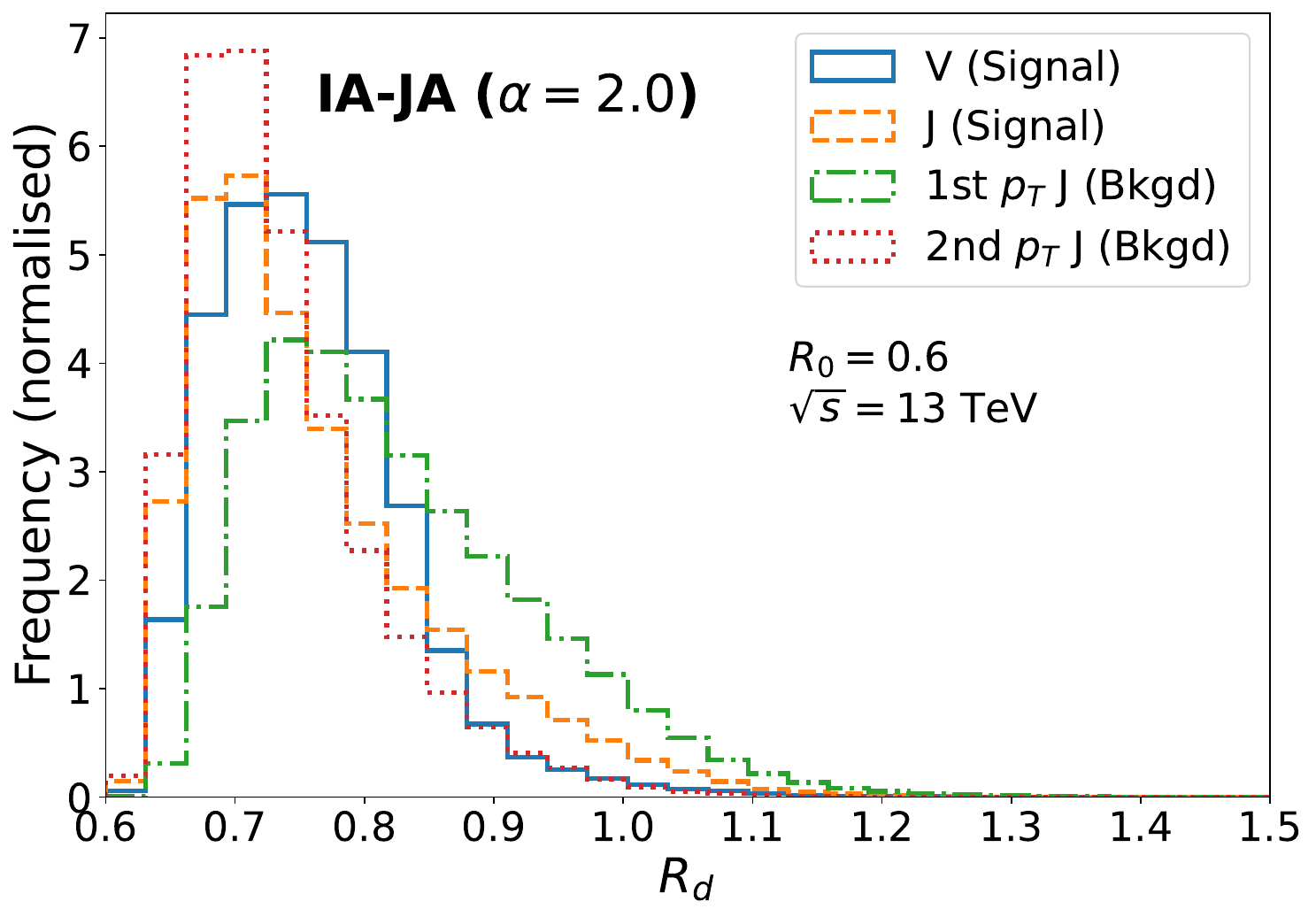}
\includegraphics[width=0.48\textwidth]{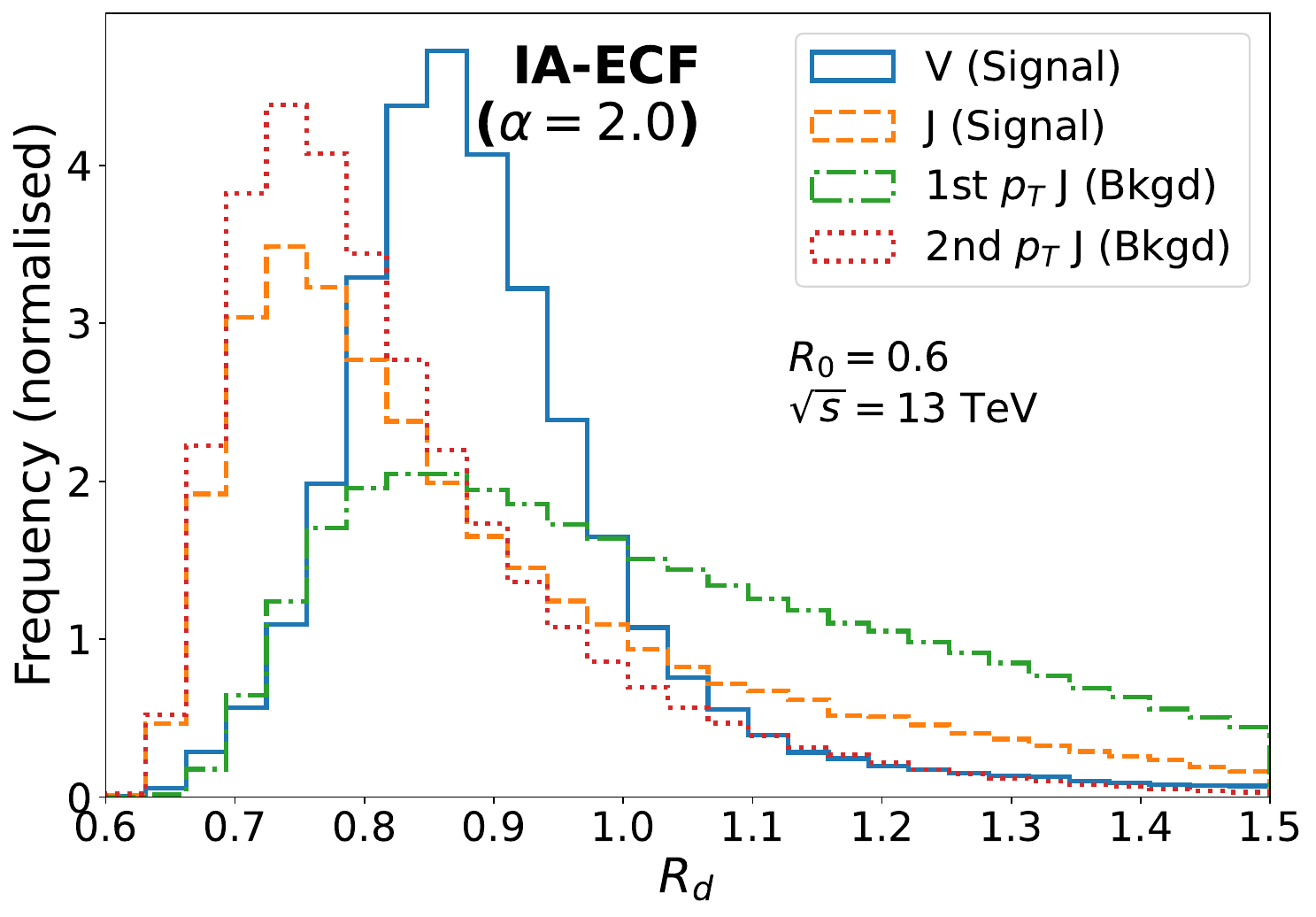}
\includegraphics[width=0.48\textwidth]{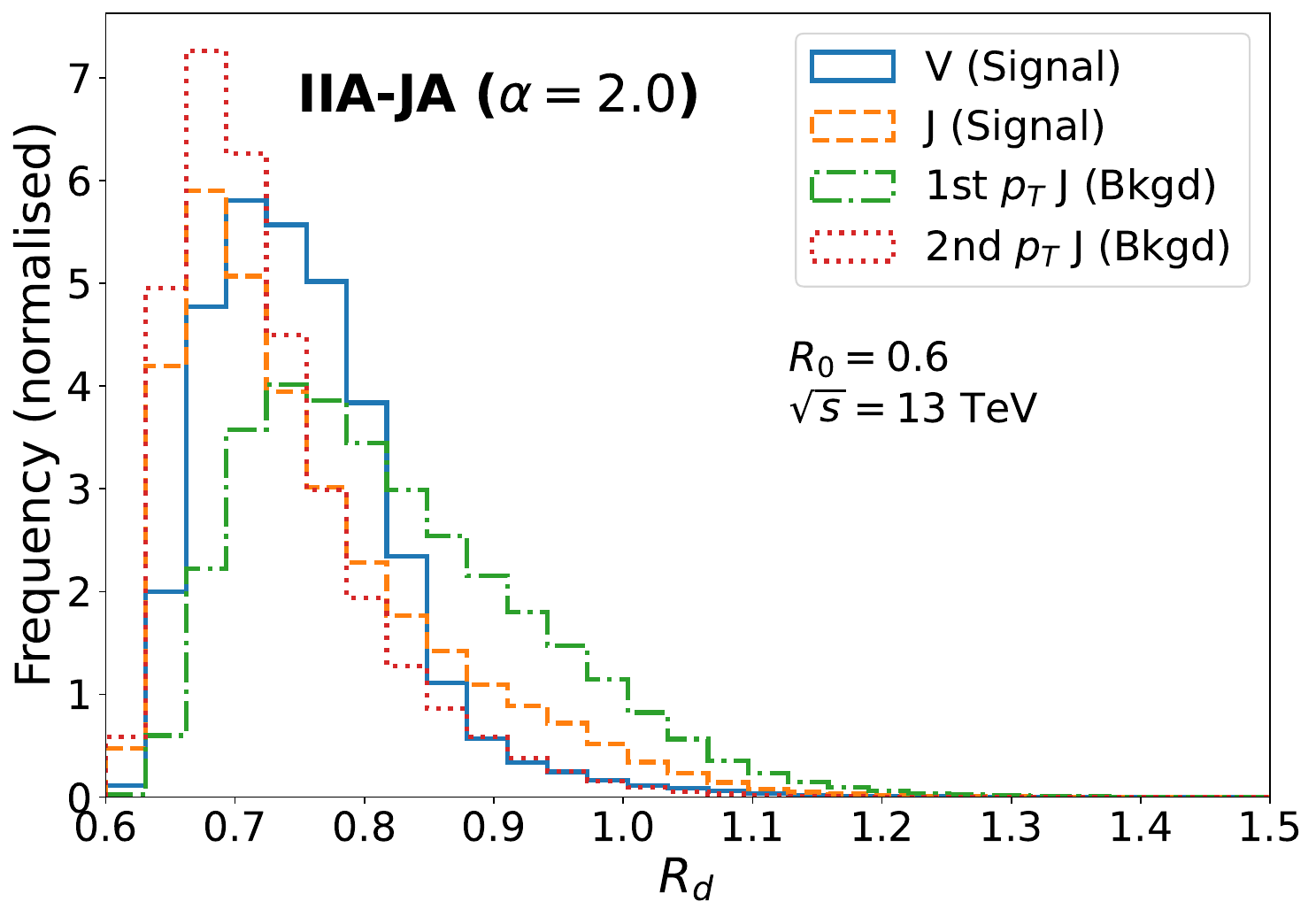}
\includegraphics[width=0.48\textwidth]{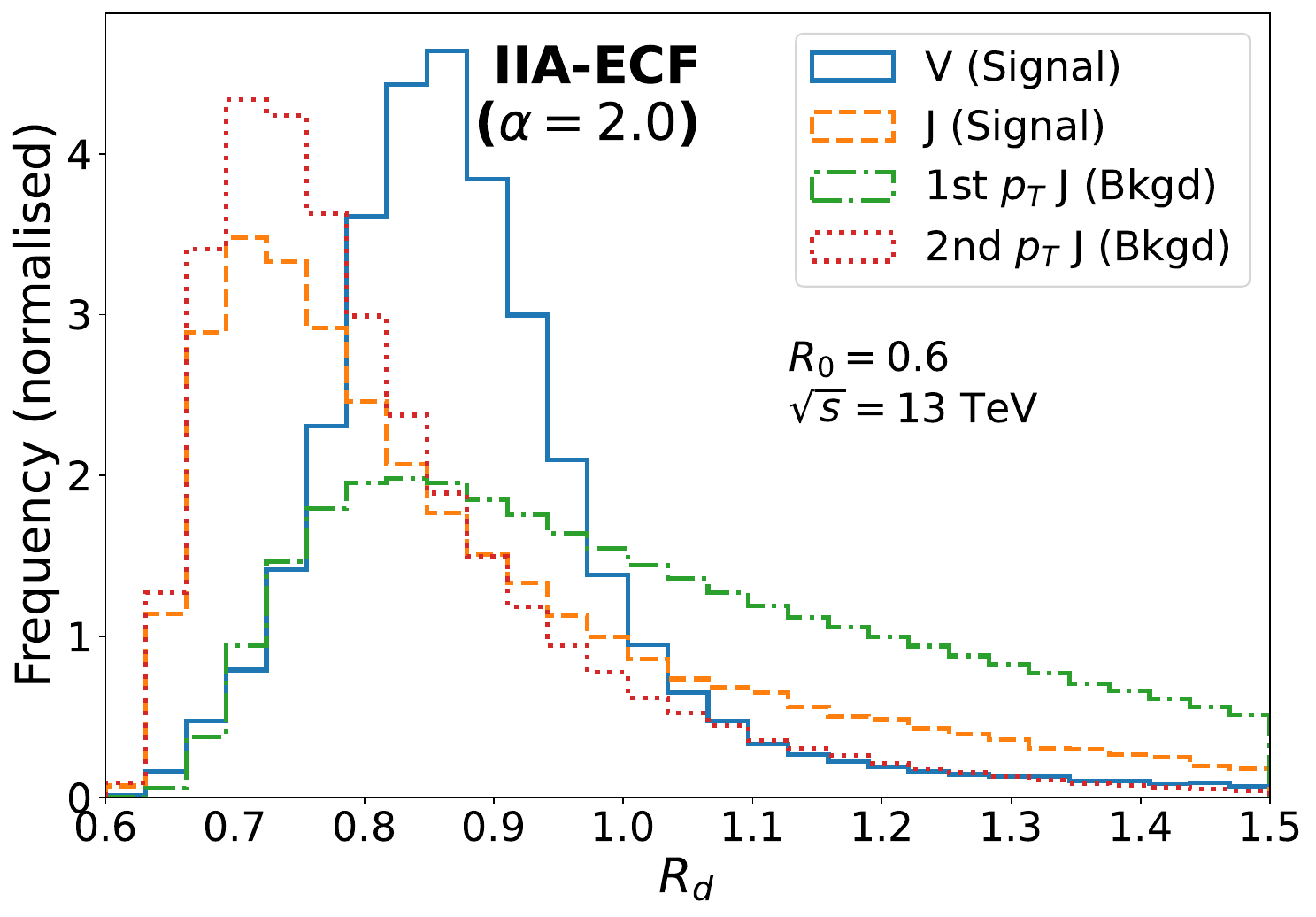}
\end{center}
\vspace{-24pt}
\caption{Normalised distributions of $R_d$ for different DR variants for the $\Vjfh$ (signal) and $\jjfh$ (background) samples with starting radius $R_0=0.6$.
Each panel shows a different DR variant, as indicated in the panel, with the variants described in Sec.~\ref{sec:drAlgo}. ECF and JA denote energy correlation functions and jet angularities, respectively.}
\label{fig:Rd-Vj500-jj}
\end{figure}

We next consider the $\Vjfh$ sample, corresponding to a more boosted kinematic regime with $p_T^V>500~\mathrm{GeV}$. The distributions of the dynamically adjusted jet radii are shown in Figure~\ref{fig:Rd-Vj500-jj}. Similar to the $\Vjth$ case, the boosted vector-boson jets preferentially acquire larger effective radii, while the accompanying light jet and the jets in the background $\jjfh$ sample are reconstructed with smaller effective radii.

\begin{figure}[!b]
\begin{center}
\includegraphics[width=0.48\textwidth]{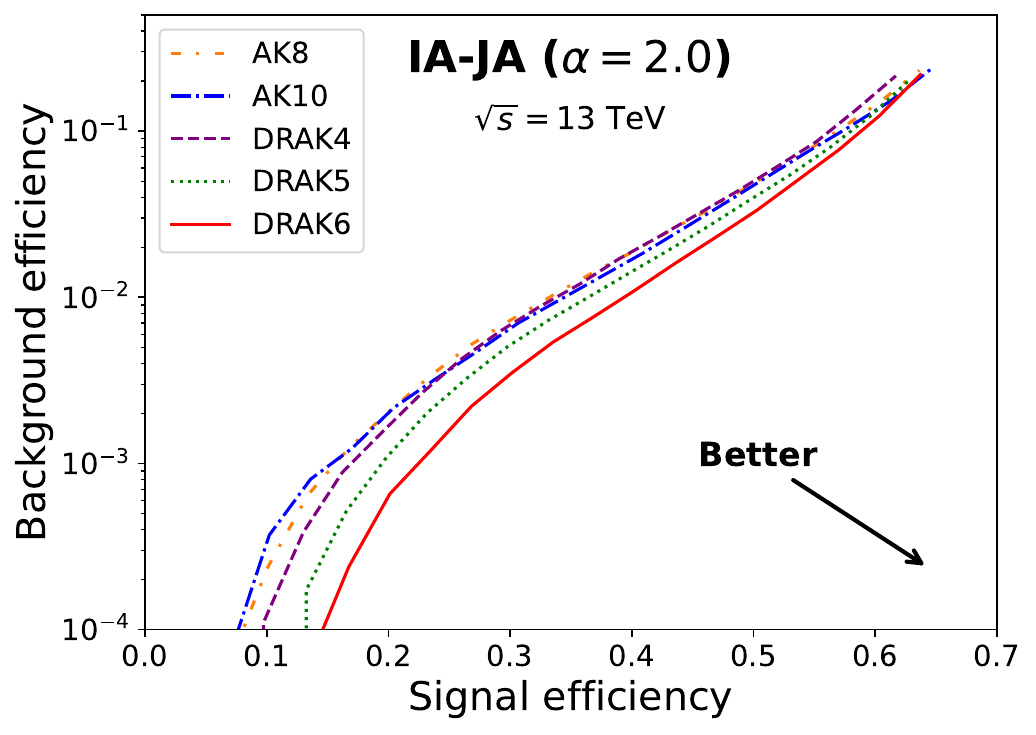}
\includegraphics[width=0.48\textwidth]{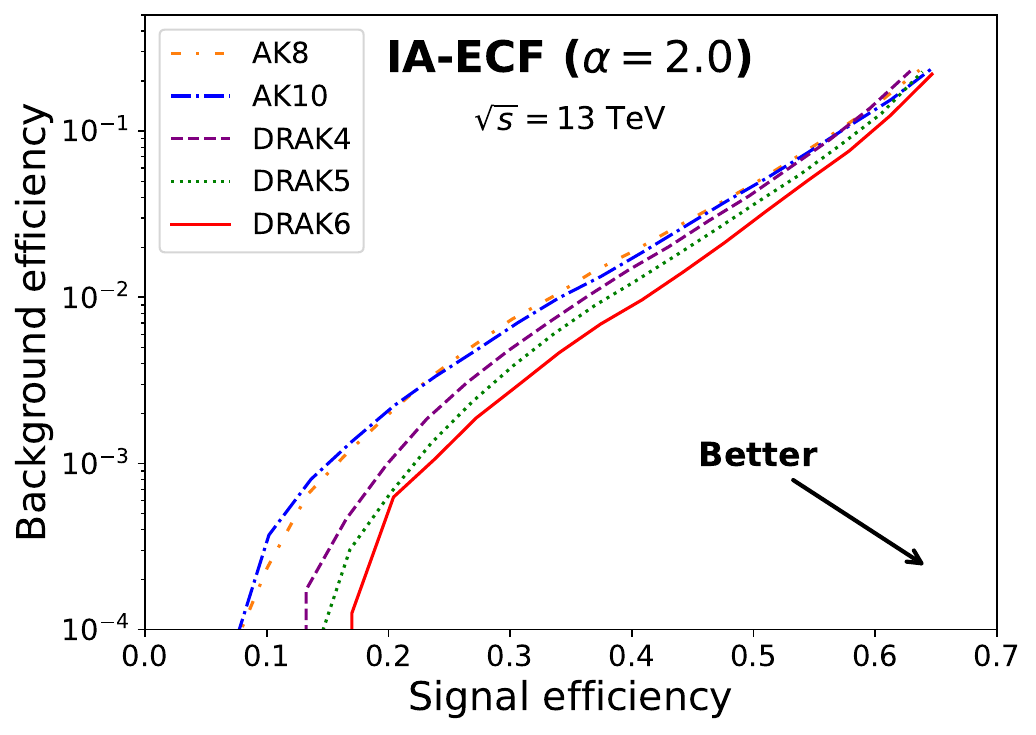}
\includegraphics[width=0.48\textwidth]{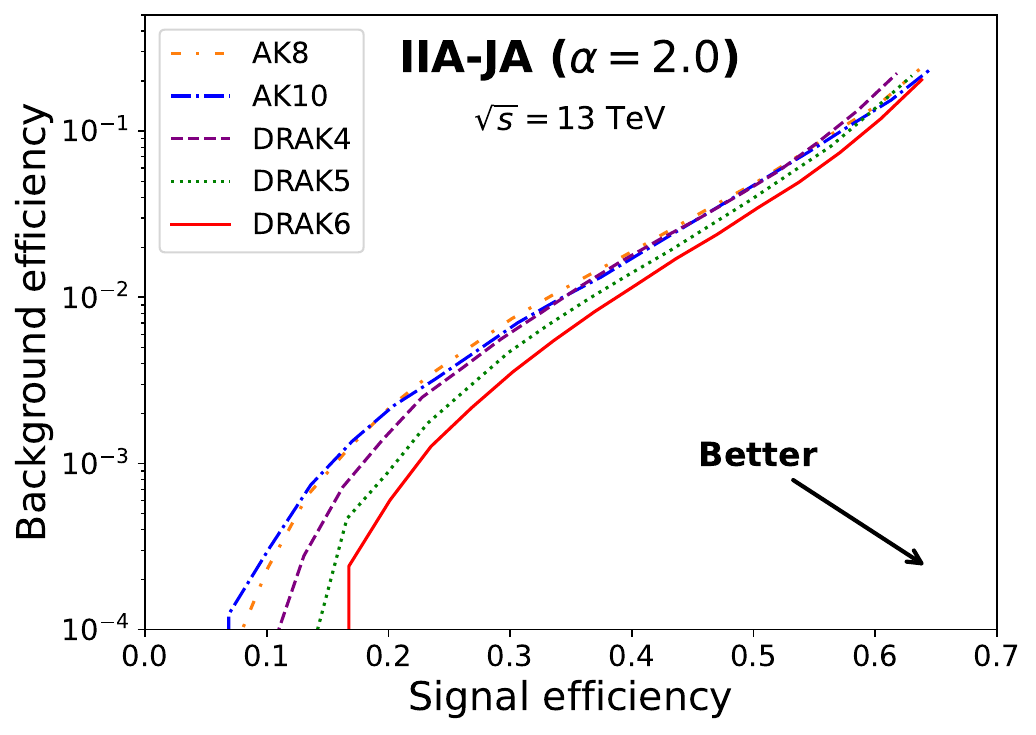}
\includegraphics[width=0.48\textwidth]{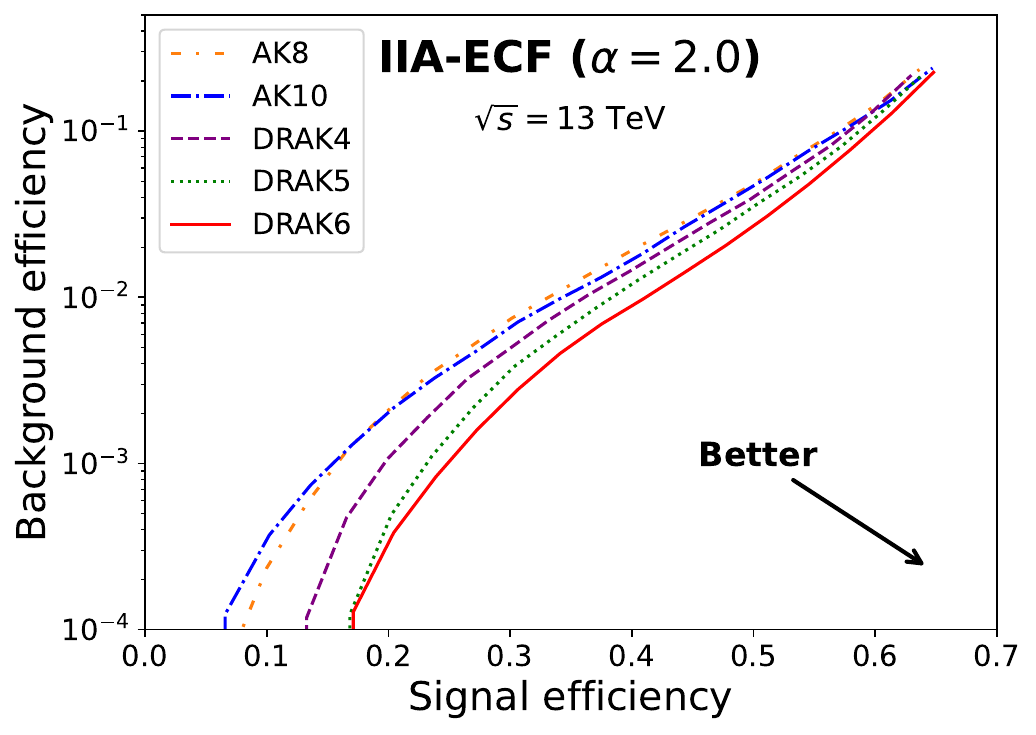}
\end{center}
\vspace{-20pt}
\caption{Comparison of ROC curves for different clustering methods for \Vjfh (signal) vs.~\jjfh (background) samples. AK8 and AK10 correspond to fixed radius anti-$k_t$ algorithms with $R=0.8$ and 1.0, respectively, while DRAK4, DRAK5, and DRAK6 correspond to dynamic radius anti-$k_t$ algorithms with starting radius $R_0=0.4$, 0.5, and 0.6, respectively. Each panel corresponds to one DR variant, as indicated in the panel, with the variants described in Sec.~\ref{sec:drAlgo}. ECF and JA denote energy correlation functions and jet angularities, respectively.}
\label{fig:Vj500-jj}
\end{figure}

The ROC curves obtained from the multivariate analysis are presented in Figure~\ref{fig:Vj500-jj}. The same set of input observables described in Sec.~\ref{sec:bdt} is used. The dynamic radius jets again demonstrate improved discrimination power compared to the corresponding fixed radius anti-$k_t$ jets. Comparing with the $\Vjth$ sample, the improvement is particularly pronounced in this highly boosted regime, where the event-by-event adaptation of the jet radius allows the decay products of the boosted vector boson to be reconstructed more effectively.
Furthermore, as in the $\Vjth$ vs.~$\jjth$ case (Sec.~\ref{sec:Vj300-jj300}), the variants based on the ECF-inspired radius modifier show better performance than those based on the JA-inspired modifier. Similarly, the type IIA construction performs better than type IA across the different radius modifiers. Overall, the IIA-ECF variant again provides the best performance among the four variants considered.

\subsection{Analysis of $\tjfh$ vs.~\jjfh Samples}
\label{sec:tj500-jj500}
\begin{figure}[!h]
\begin{center}
\includegraphics[width=0.48\textwidth]{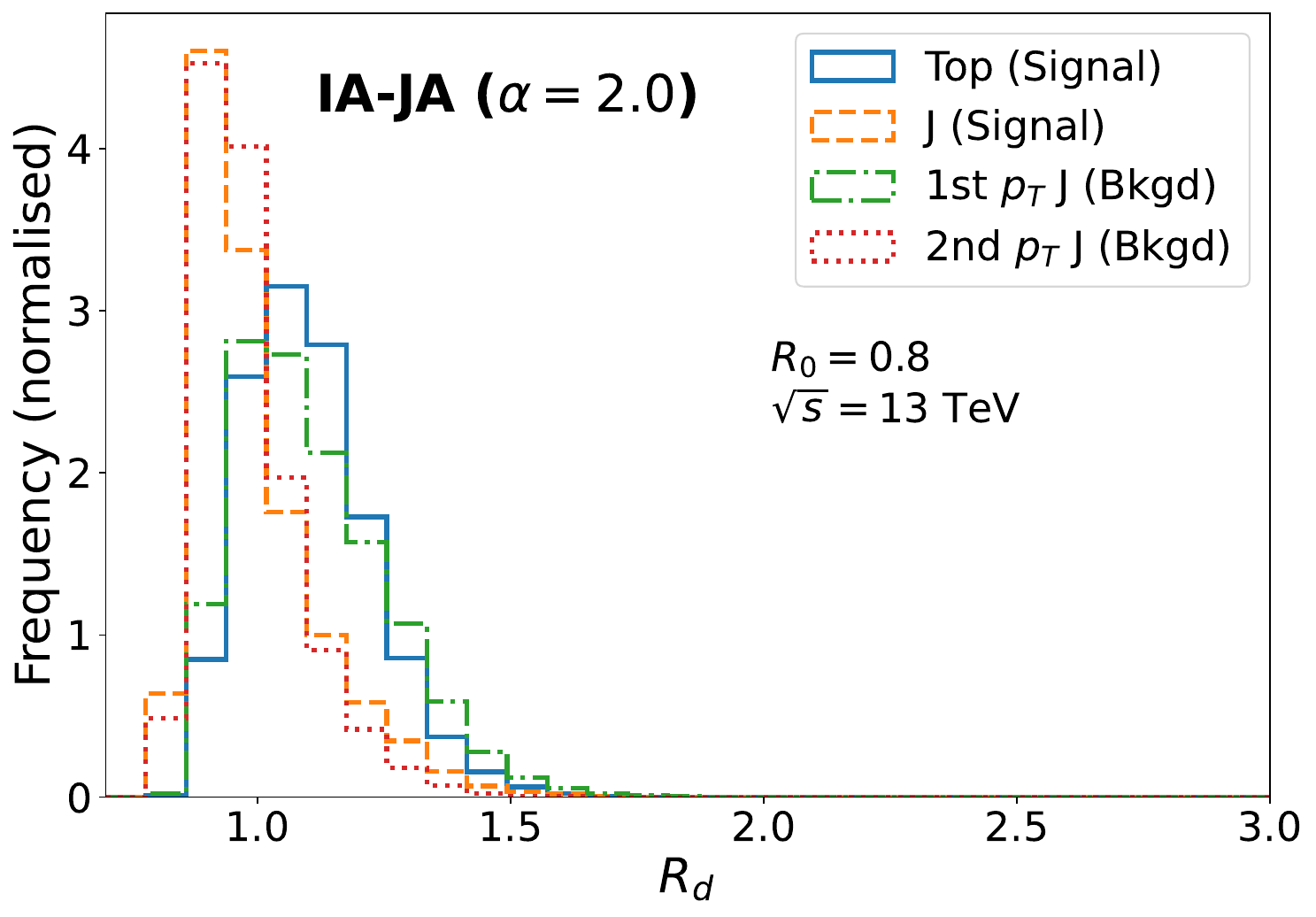}
\includegraphics[width=0.48\textwidth]{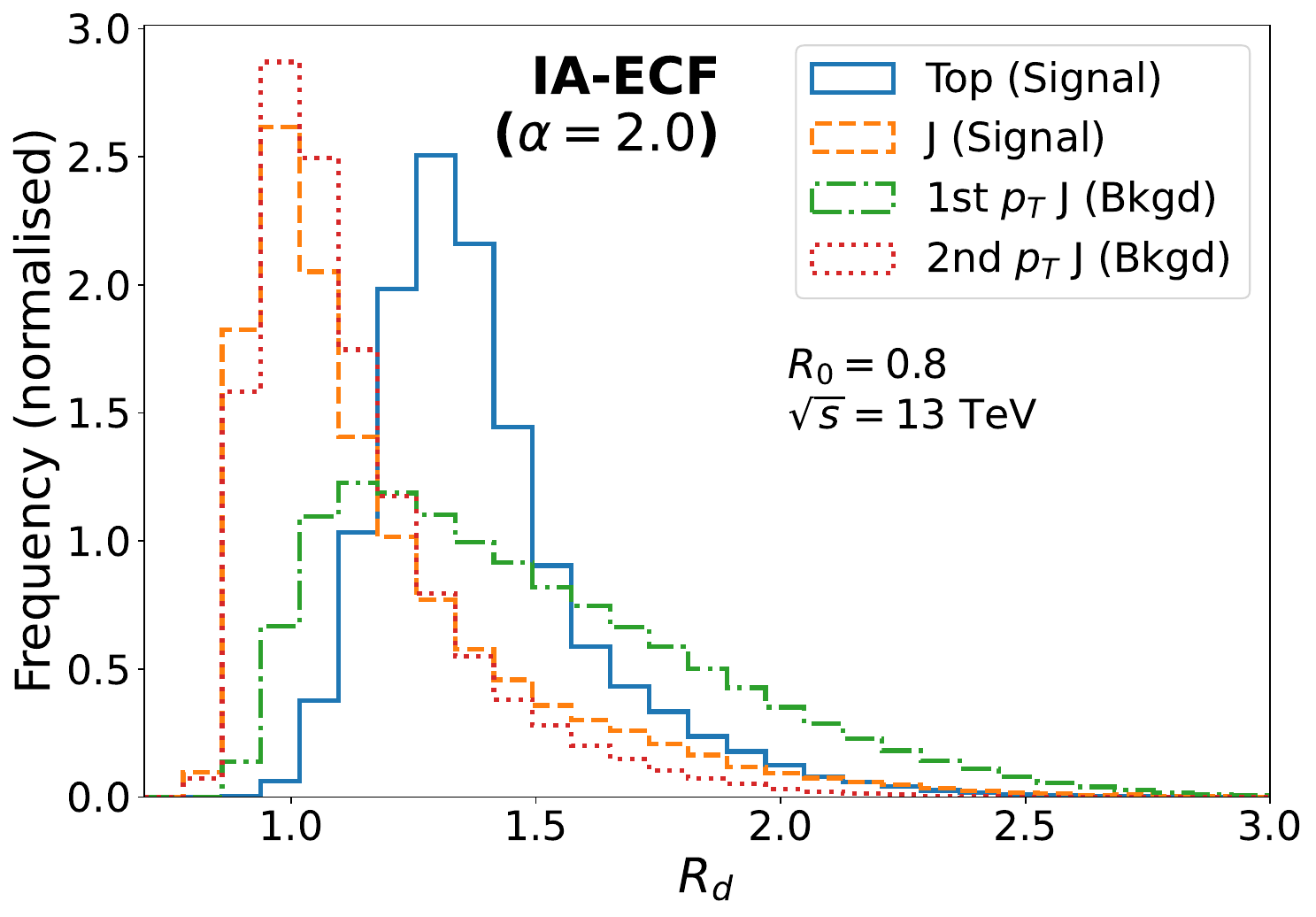}
\includegraphics[width=0.48\textwidth]{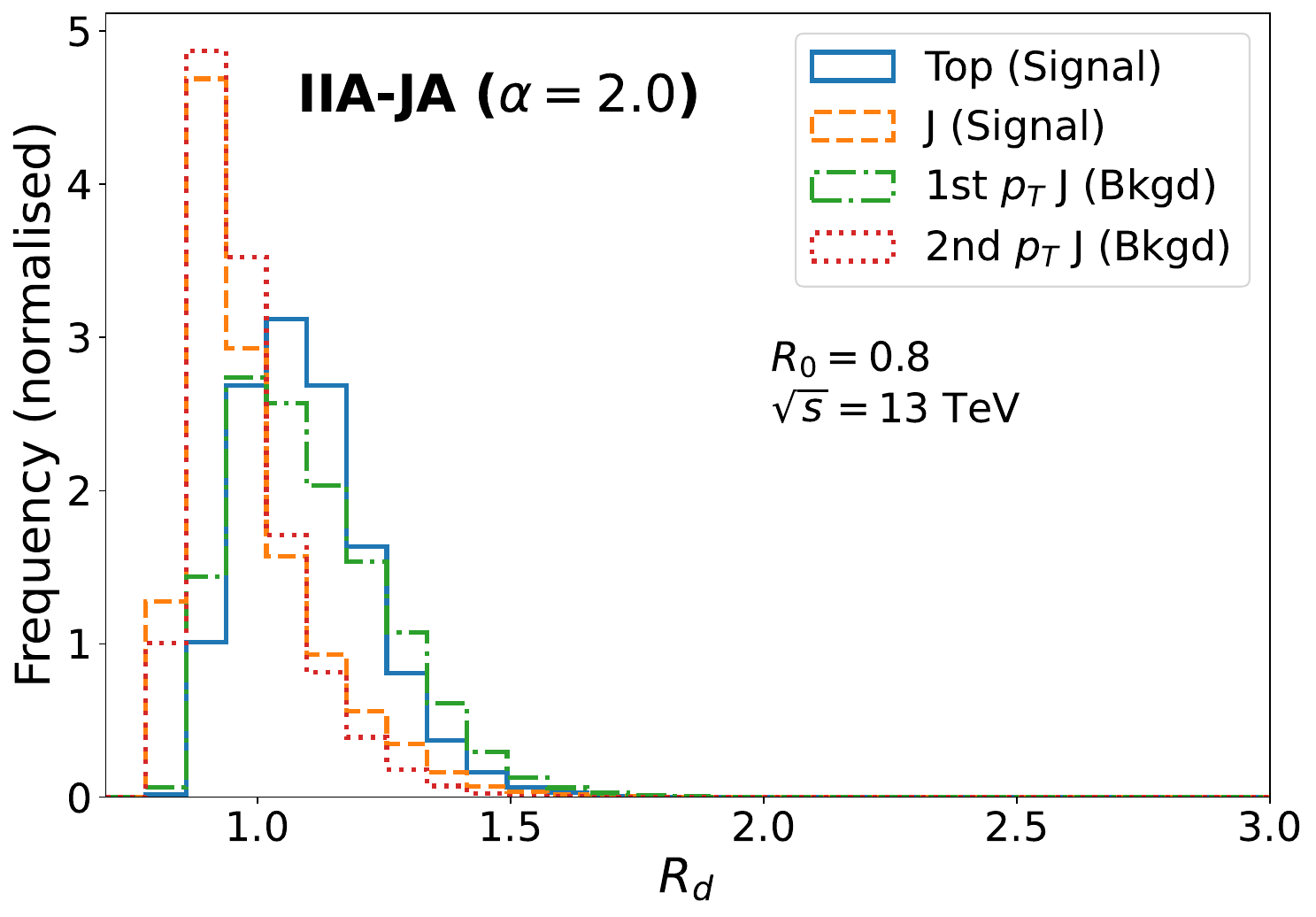}
\includegraphics[width=0.48\textwidth]{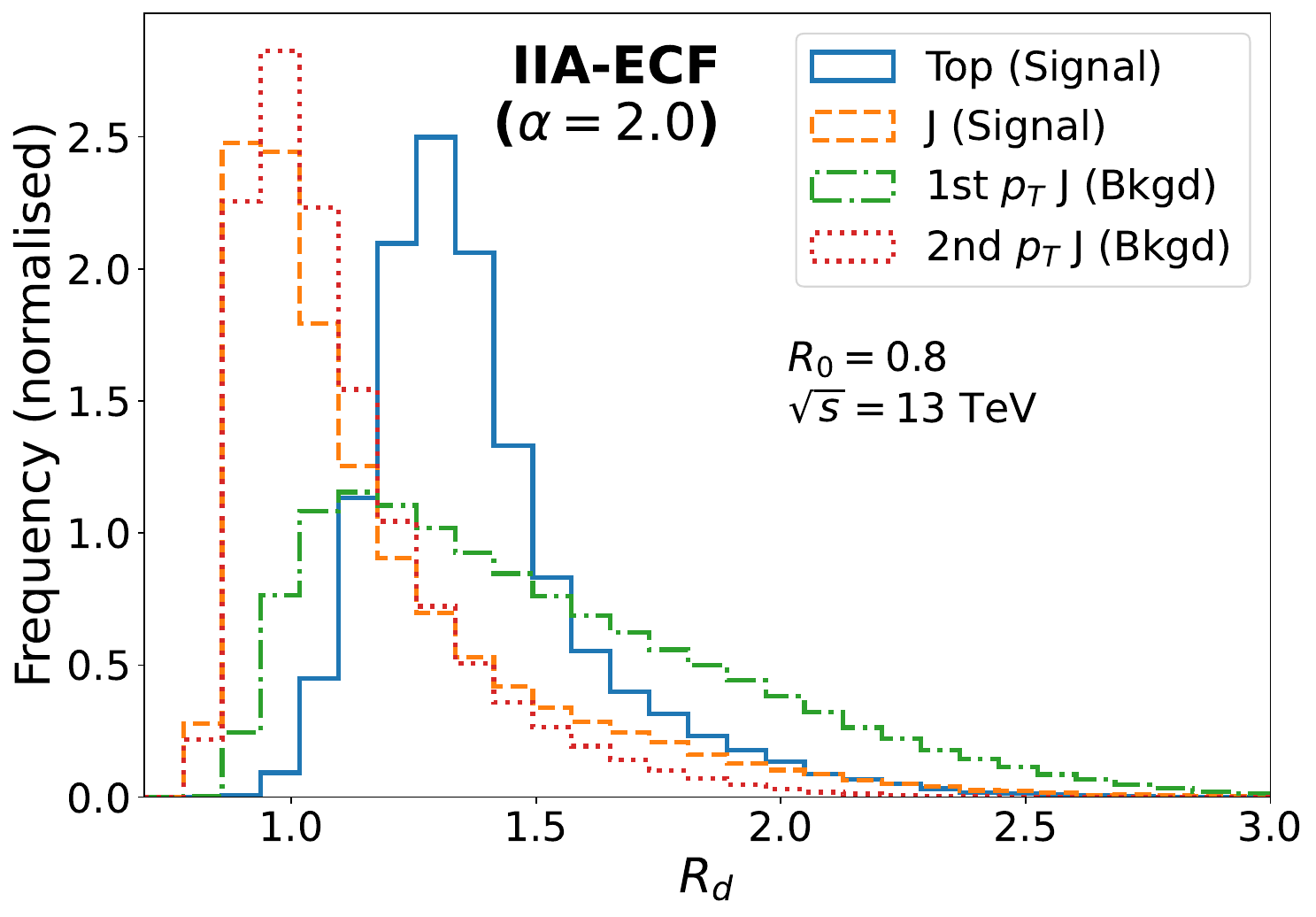}
\end{center}
\caption{Normalised distributions of $R_d$ for different DR variants for the $\tjfh$ (signal) and $\jjfh$ (background) samples with starting radius $R_0=0.8$.
Each panel shows a different DR variant, as indicated in the panel, with the variants described in Sec.~\ref{sec:drAlgo}. ECF and JA denote energy correlation functions and jet angularities, respectively.}
\label{fig:Rd-tj-jj}
\end{figure}

Finally, we consider the $\tjfh$ sample, requiring a boosted top quark with $p_T^t>500~\mathrm{GeV}$ at the parton-level event generation. The distributions of the dynamically adjusted radii for the reconstructed top jet and the accompanying light jet are shown in Figure~\ref{fig:Rd-tj-jj}. Due to the larger angular extent of the three-prong top-quark decay system, the reconstructed top jets typically acquire larger effective radii than the accompanying light jets and the jets in the QCD background sample.

\begin{figure}[h]
\begin{center}
\includegraphics[width=0.48\textwidth]{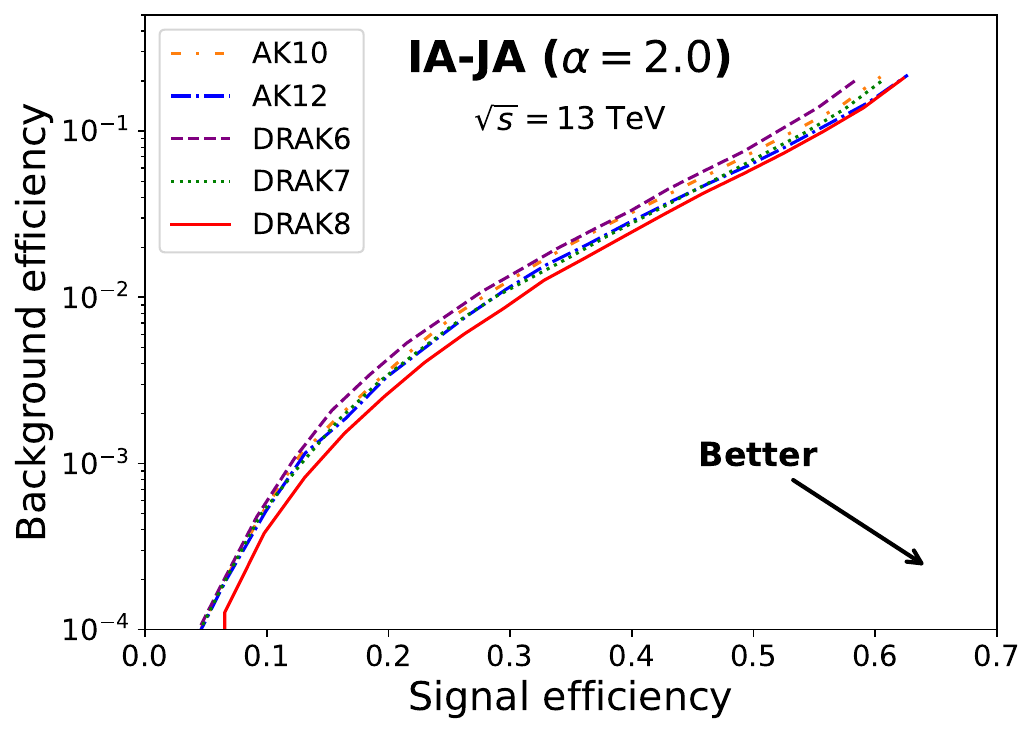}
\includegraphics[width=0.48\textwidth]{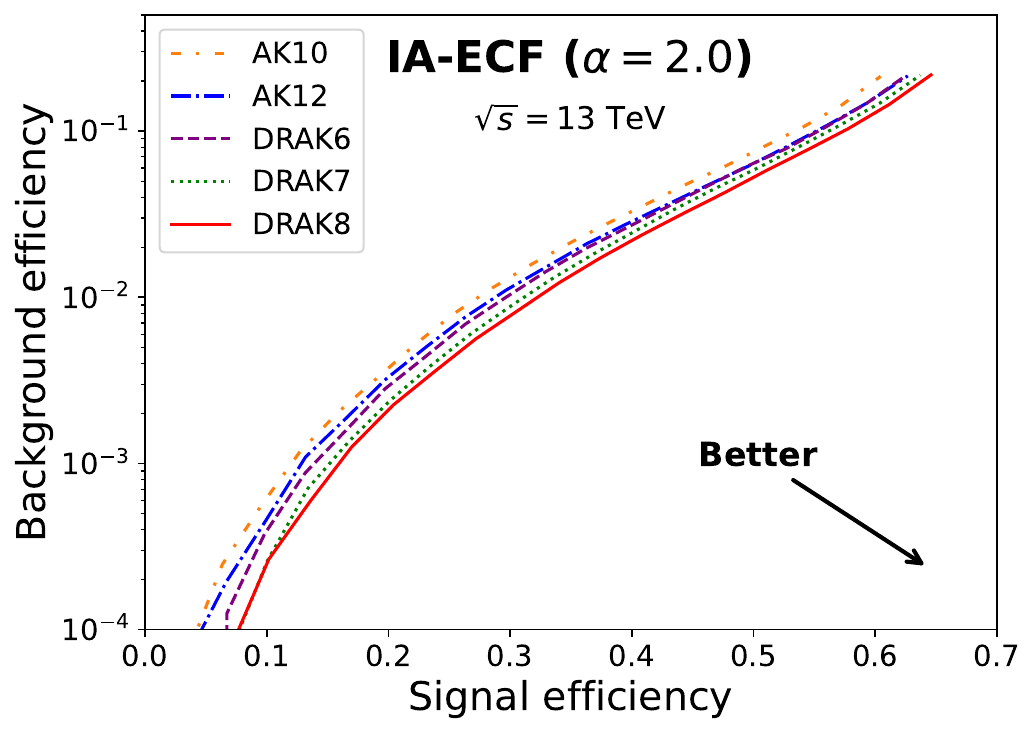}
\includegraphics[width=0.48\textwidth]{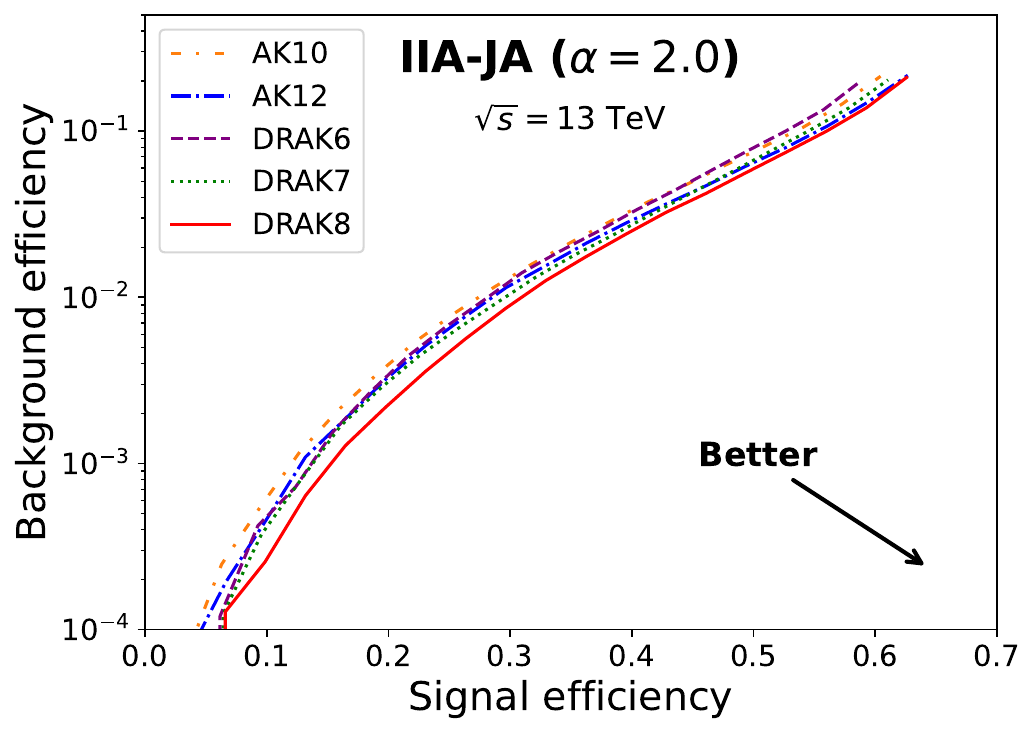}
\includegraphics[width=0.48\textwidth]{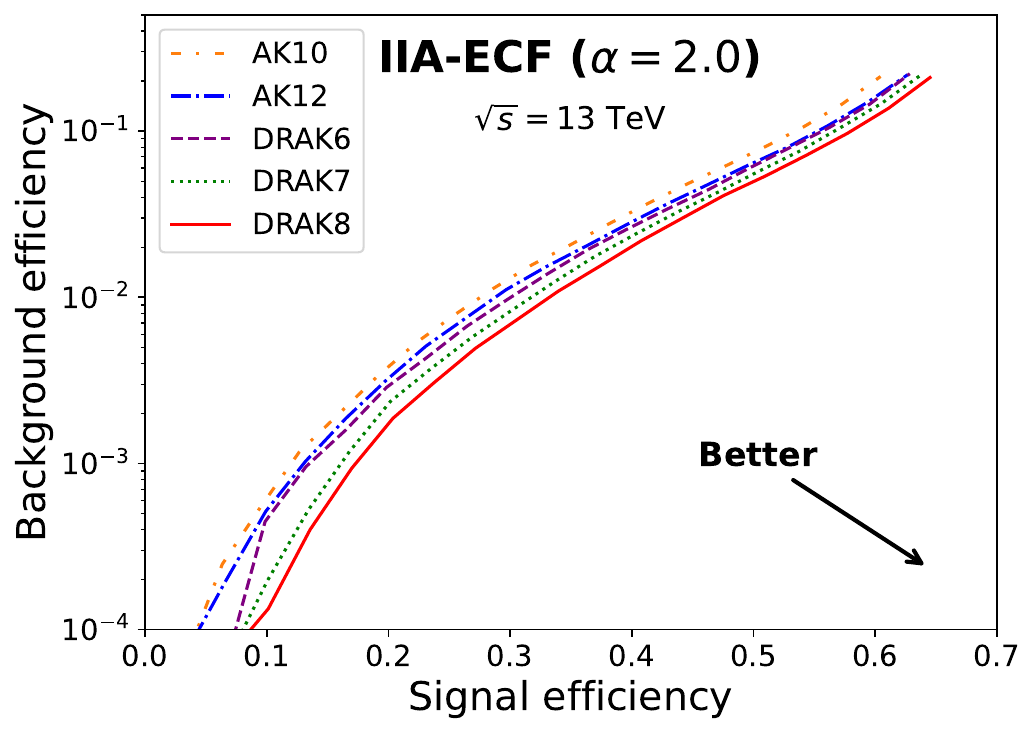}
\end{center}
\caption{Comparison of ROC curves for different clustering methods for \tjfh (signal) vs.~\jjfh (background) samples. AK10 and AK12 correspond to fixed radius anti-$k_t$ algorithms with $R=1.0$ and 1.2, respectively, while DRAK6, DRAK7, and DRAK8 correspond to dynamic radius anti-$k_t$ algorithms with starting radius $R_0=0.6$, 0.7, and 0.8, respectively. Each panel corresponds to one DR variant, as indicated in the panel, with the variants described in Sec.~\ref{sec:drAlgo}. ECF and JA denote energy correlation functions and jet angularities, respectively.}
\label{fig:tj-jj}
\end{figure}

The performance of the multivariate analysis is summarised by the ROC curves shown in Figure~\ref{fig:tj-jj}. As in the vector-boson analyses, the dynamic radius approach provides improved signal-background separation compared to fixed radius anti-$k_t$ clustering. These results indicate that incorporating information on the JSS into the determination of the jet radius can provide a more adaptive reconstruction of boosted heavy-particle decays and consequently improve the discrimination between signal and background.
A comparison among four panels of Figure~\ref{fig:tj-jj} shows a similar trend to that observed for the $\Vjth$ vs.~$\jjth$ (Sec.~\ref{sec:Vj300-jj300}) and $\Vjfh$ vs.~$\jjfh$ (Sec.~\ref{sec:Vj500-jj500}) samples. The ECF-inspired radius modifiers perform better than the corresponding JA-inspired modifiers, while the type IIA construction performs better than type IA. Among the four variants considered, the IIA-ECF variant provides the best overall performance.

\section{Summary and Outlook}
\label{sec:summary}
Jets are among the most central physics objects in high-energy collider experiments and play a crucial role in both SM precision measurements and searches for BSM physics. At hadron colliders such as the LHC, jets arise from energetic quarks and gluons undergoing QCD radiation and hadronisation as well as from the hadronic decays of boosted heavy particles. Therefore, their efficient reconstruction and characterisation are essential for understanding the underlying event topology. Over the years, sequential recombination algorithms such as the $k_t$, anti-$k_t$, and Cambridge/Aachen algorithms have become the standard tools for jet reconstruction. These algorithms employ a fixed jet radius parameter, which determines the angular extent of the reconstructed jet in the rapidity-azimuth plane. Although highly successful, the use of a fixed radius introduces intrinsic limitations, particularly in scenarios involving jets originating from different physical processes or produced in different boosted regimes.

To overcome the inadequacy of traditional fixed radius approaches, the dynamic radius jet clustering algorithm was introduced in Ref.\,\cite{Mukhopadhyaya:2023rsb}. The central idea of the DR framework is to allow the jet radius to evolve adaptively during the clustering procedure, depending on the local kinematic and substructure properties of the constituents within an evolving proto-jet. In this approach, clustering starts from an initial seed radius $R_0$, and the effective jet radius dynamically grows with the help of a radius modifier.
Such an adaptive framework enables the algorithm to naturally accommodate varying jet sizes and has been shown to better encapsulate jets originating from different boosted regimes as well as different physical origins.

In the previous formulation of the DR jet algorithm, a specific choice of radius modifier was employed, namely the $p_T$-weighted standard deviation of inter-constituent distances in the rapidity-azimuth plane. Using this particular modifier, previous studies demonstrated the potential usefulness of the dynamic-$R$ approach in collider searches, especially in boosted BSM scenarios, where it showed improved performance over its fixed-$R$ counterparts through detailed simulation studies\,\cite{Kar:2022hxn,Mukhopadhyaya:2023akv,Ghosh:2025gue,Ghosh:2025gdq}. These observations strongly motivate a broader and more systematic exploration of the DR framework.

In this work, we present a generalised framework based on jet substructure observables such as jet angularities and energy correlation functions of the dynamic radius jet clustering algorithm by extending the idea beyond a specific choice of radius modifier considered in the previous implementation. The purpose of this generalisation is to explore a wider class of adaptive jet-clustering strategies and investigate their applicability across different physical scenarios and kinematic regimes. Different variants of the DR approach are expected to be useful for different jet topologies, including both narrow and fat jets, as well as moderately and extremely boosted regimes. A comprehensive phenomenological study has been performed incorporating multiple variants of the DR algorithm along with realistic detector and pileup effects.

For the analysis, we consider several SM benchmark samples relevant for boosted jet studies at the 14 TeV HL-LHC. Signal samples include $V$+jet with $p_T^V > 300$ GeV corresponding to moderately boosted configurations, highly boosted $V$+jet with $p_T^V > 500$ GeV, and top+jet with $p_T^t > 500$ GeV representing both wide-jet and narrow-jet configurations. Corresponding QCD dijet background samples were generated for $p_T > 300$ GeV and $p_T > 500$ GeV regions. Event generation was performed using MadGraph5\_aMC\@NLO, followed by parton showering and hadronisation using Pythia8. To emulate realistic HL-LHC conditions, detector simulation was carried out with Delphes, including an average pileup of 150 interactions. Pileup mitigation was subsequently performed using the PUPPI algorithm. Various JSS observables and techniques, including $N$-subjettiness,  energy correlation functions and related discriminating variables, were employed to study the tagging performance of the reconstructed jets.

After pileup mitigation through the PUPPI framework, the performance of the algorithms was studied using multivariate analysis techniques based on XGBoost. Using ROC-based analyses, a systematic and comprehensive comparison was performed between the best-performing fixed radius anti-$k_t$ configuration and the best-performing dynamic radius anti-$k_t$ variants. The study indicates that the optimal DR version consistently outperforms the corresponding best fixed radius anti-$k_t$ configuration across three benchmark scenarios considered. Furthermore, the generalised DR approaches exhibit good robustness against pileup contamination once PUPPI mitigation is applied, indicating that the adaptive nature of the algorithm does not introduce significant pileup sensitivity.

The improvement is found to be most prominent in the highly boosted regions, particularly for the $V$+jet samples, while relatively moderate improvements are observed for top-jet scenarios. These observations suggest that the DR framework is especially effective in collider environments involving highly boosted objects, which are among the primary targets of present and future SM and BSM searches at the HL-LHC. The generalised framework therefore provides a promising direction for improving jet reconstruction and tagging performance in challenging boosted environments.

Overall, this work establishes the generalised dynamic radius framework as a flexible and powerful extension of conventional jet clustering algorithms. The study opens several avenues for future exploration, including optimisation of additional radius modifiers, incorporation of machine-learning-inspired adaptive strategies within the clustering procedure, and extension to other collider environments and detector conditions. Further studies involving detector-level calibrations, systematic uncertainties, and experimental implementation strategies will also be important for assessing the full potential of the approach. The present work strongly advocates the inclusion of DR methodologies in future CMS and ATLAS analyses, particularly in boosted-object searches, and highlights its broader applicability for future collider programs beyond the HL-LHC.

\acknowledgments{The authors acknowledge the High-Performance Computing facility Nandadevi at The Institute of Mathematical Sciences for supporting the computational needs of this work. The work of S.~D.~is co-funded by the European Union's Horizon Europe research and innovation program under the Marie Sk{\l}odowska-Curie COFUND Postdoctoral Programme grant agreement No. 101081355-SMASH and by the Republic of Slovenia and the European Union from the European Regional Development Fund.}

\section*{Data Availability Statement}
The data used in this study are simulated using standard publicly available software packages, with the simulation and analysis methods described in detail in the article.


\appendix
\section{Soft Drop Jet Mass Distribution}
\label{sec:jetmass}
\begin{figure}[!h]
\begin{center}
\includegraphics[width=0.48\textwidth]{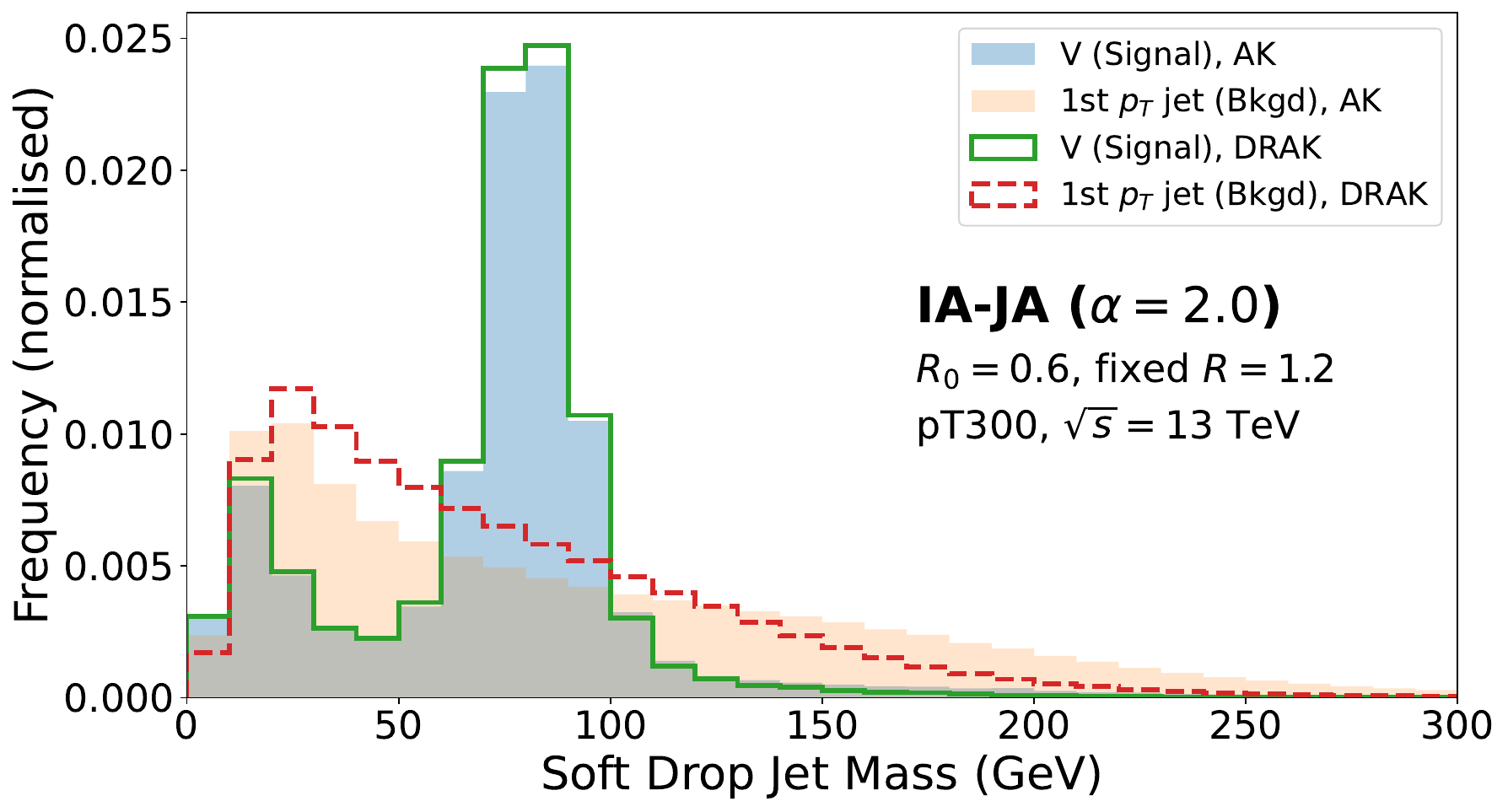}
\includegraphics[width=0.48\textwidth]{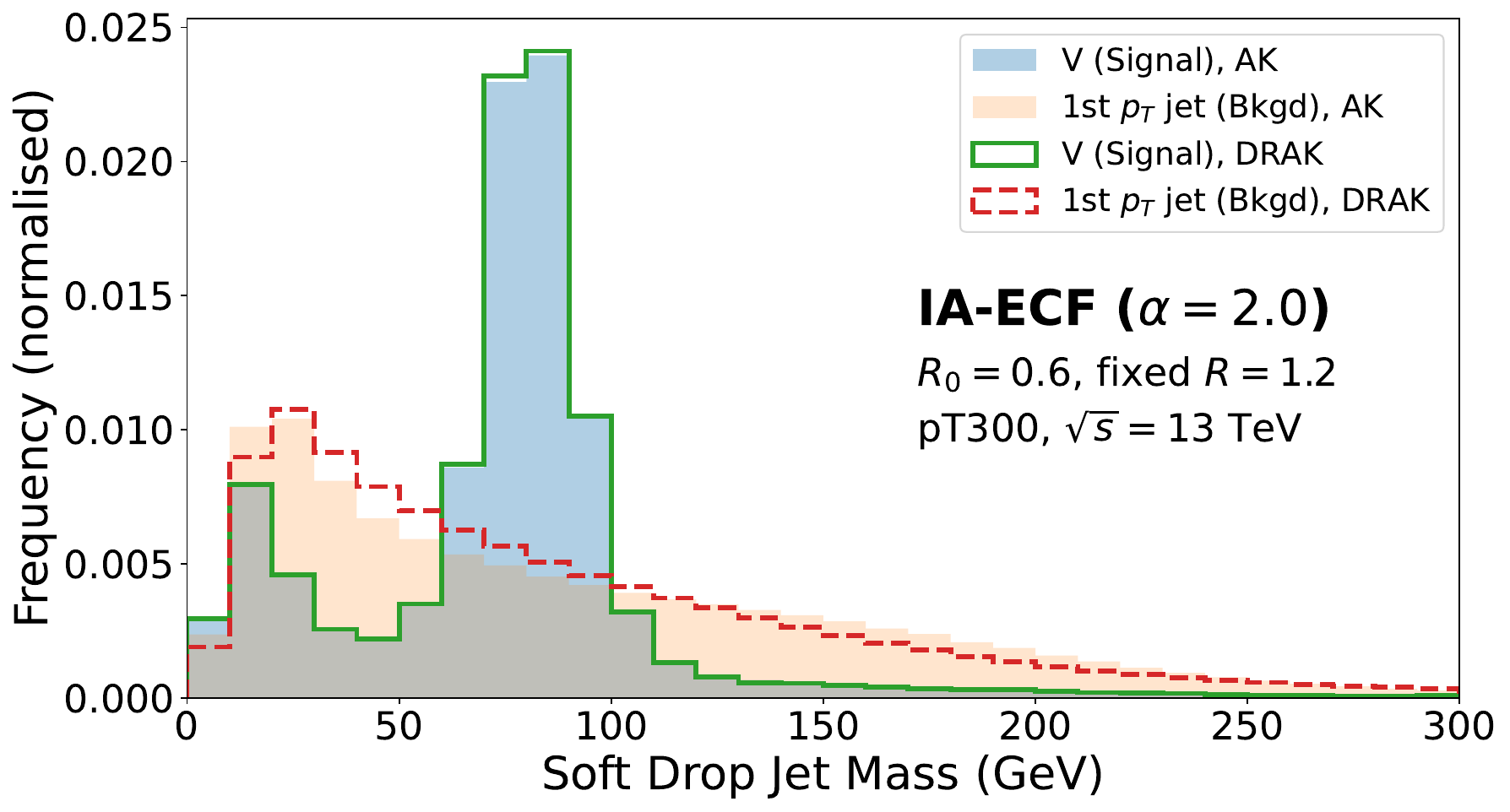}
\includegraphics[width=0.48\textwidth]{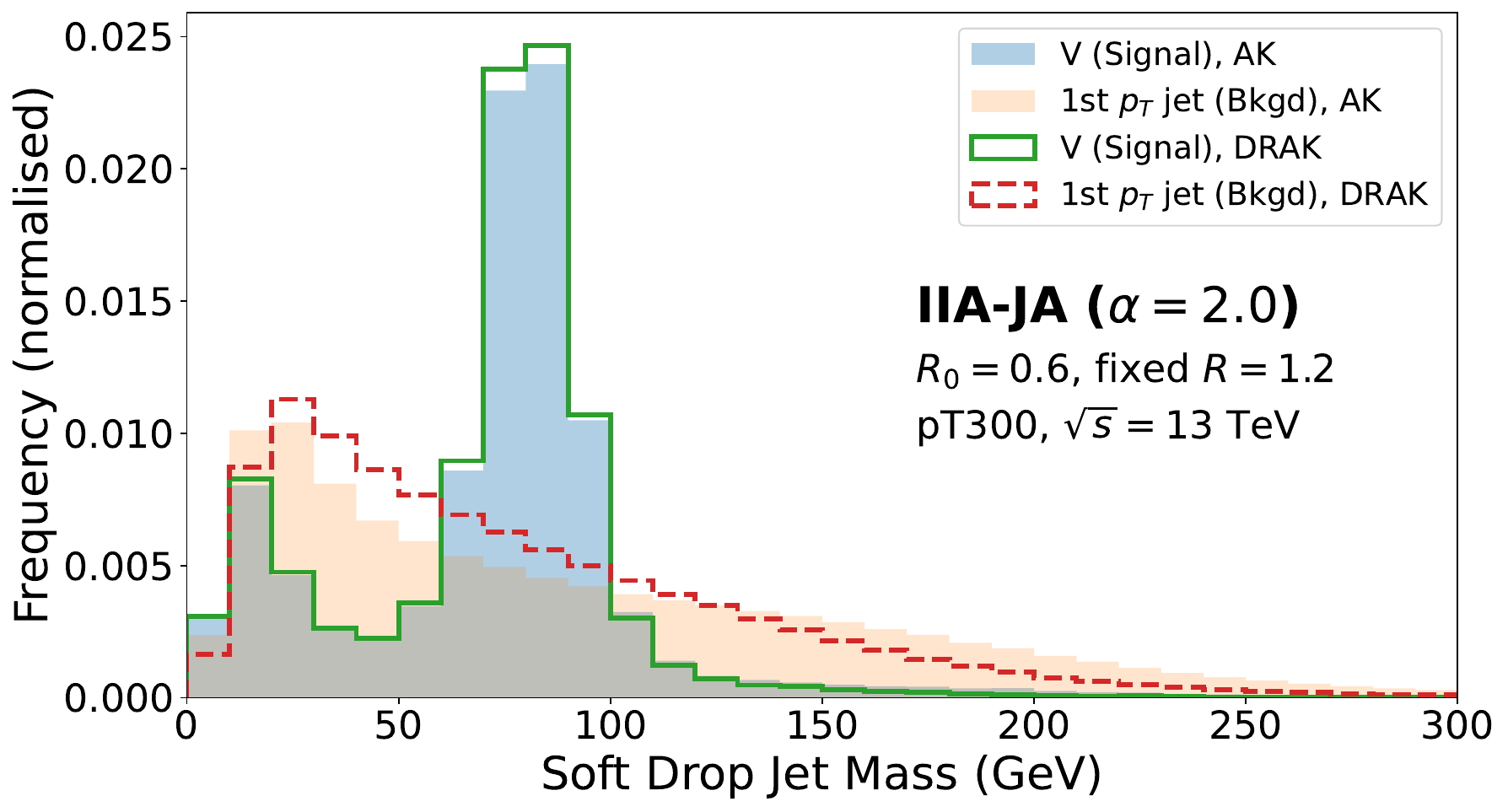}
\includegraphics[width=0.48\textwidth]{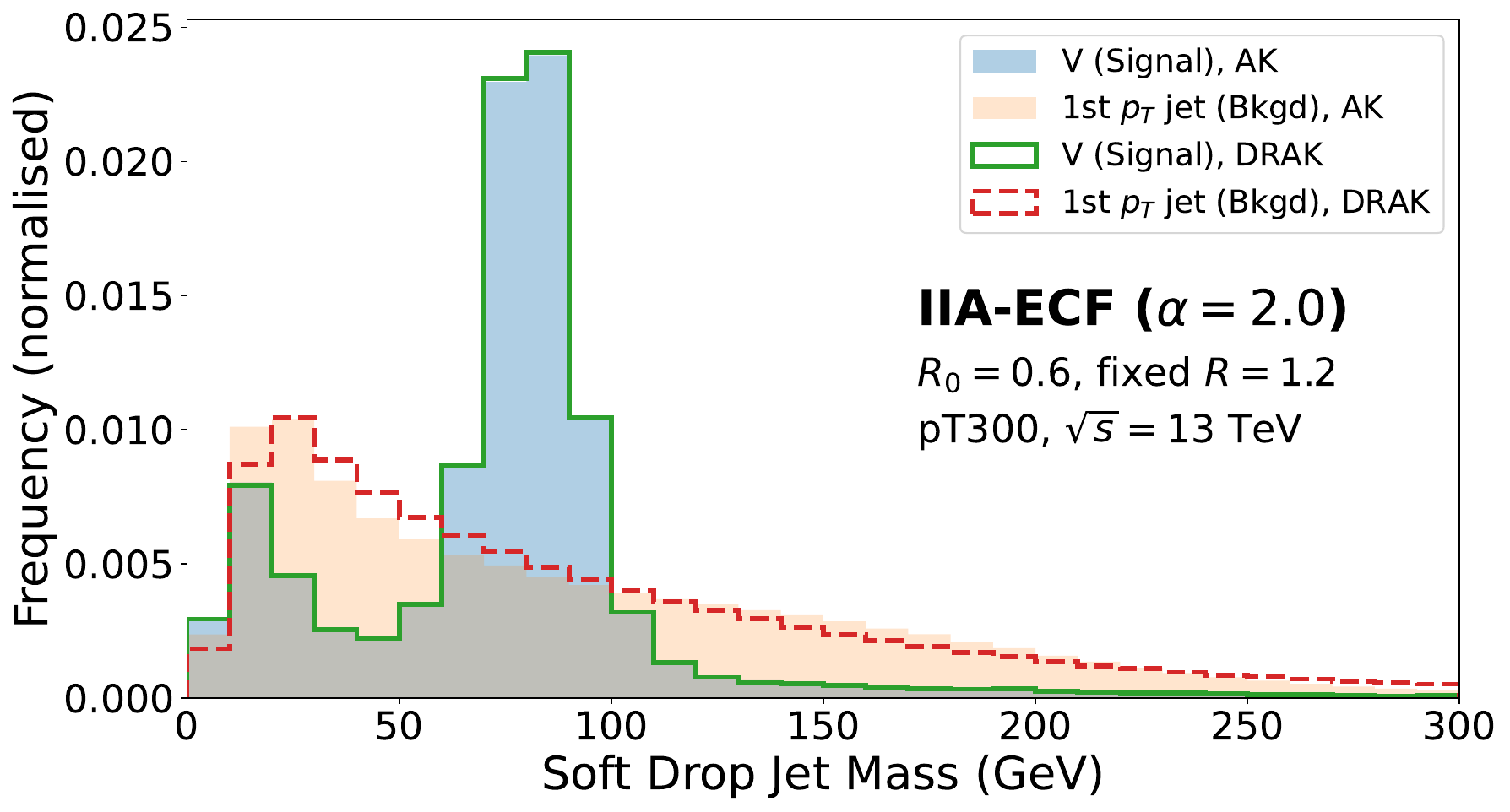}
\caption{Normalised distributions of the Soft Drop jet mass for different DR variants (unfilled histograms) for the $\Vjth$ (signal) and $\jjth$ (background) samples with starting radius $R_0=0.6$. Distributions corresponding to the fixed radius anti-$k_t$ algorithm (AK) with $R=1.0$ are also shown as a filled histogram in each panel.
Each panel shows a different DR variant, as indicated in the panel, with the variants described in Sec.~\ref{sec:drAlgo}. ECF and JA denote energy correlation functions and jet angularities, respectively.}
\label{fig:Mass-Vj300-jj}
\end{center}
\end{figure}

\begin{figure}[!h]
\begin{center}
\includegraphics[width=0.48\textwidth]{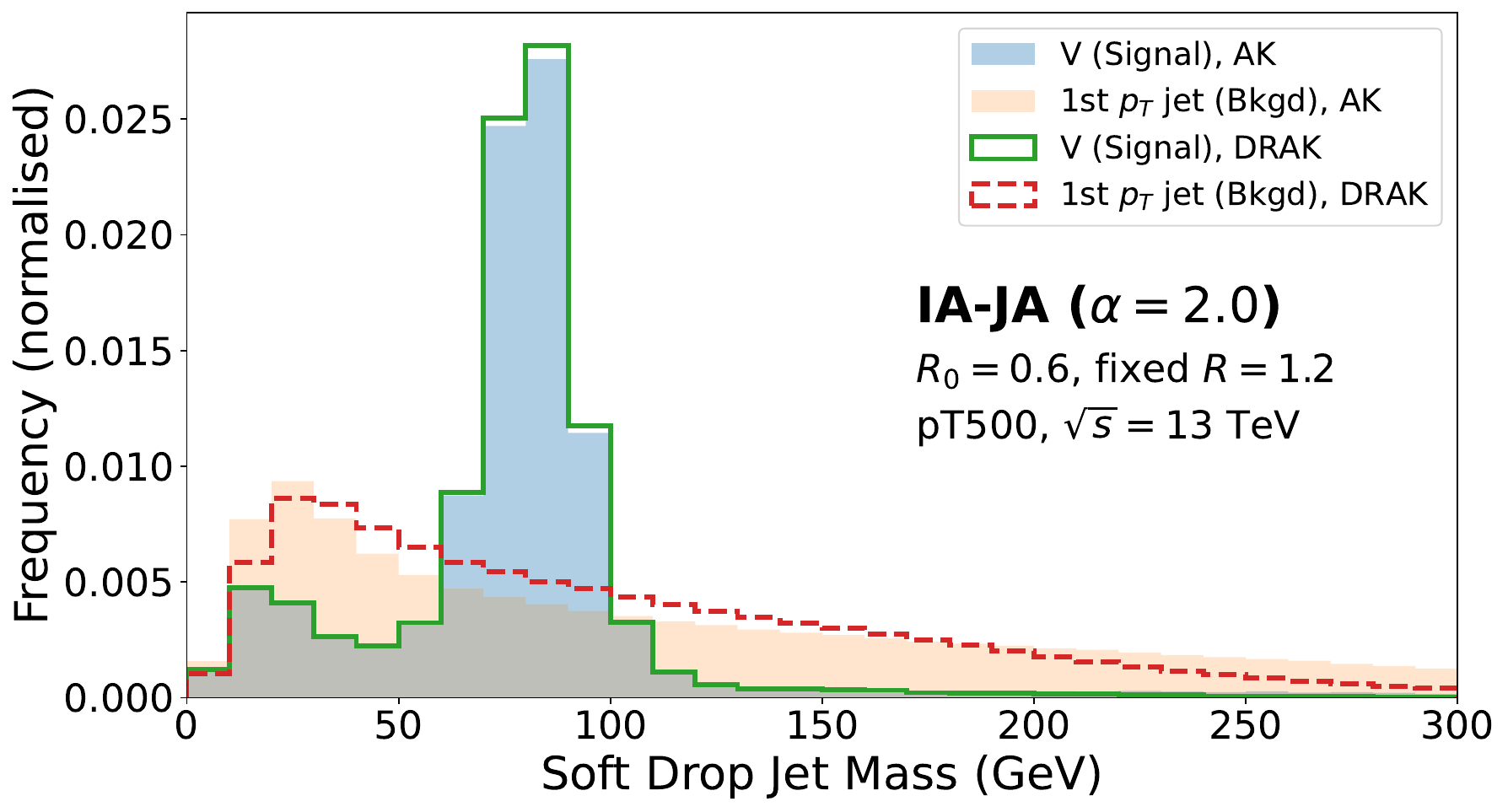}
\includegraphics[width=0.48\textwidth]{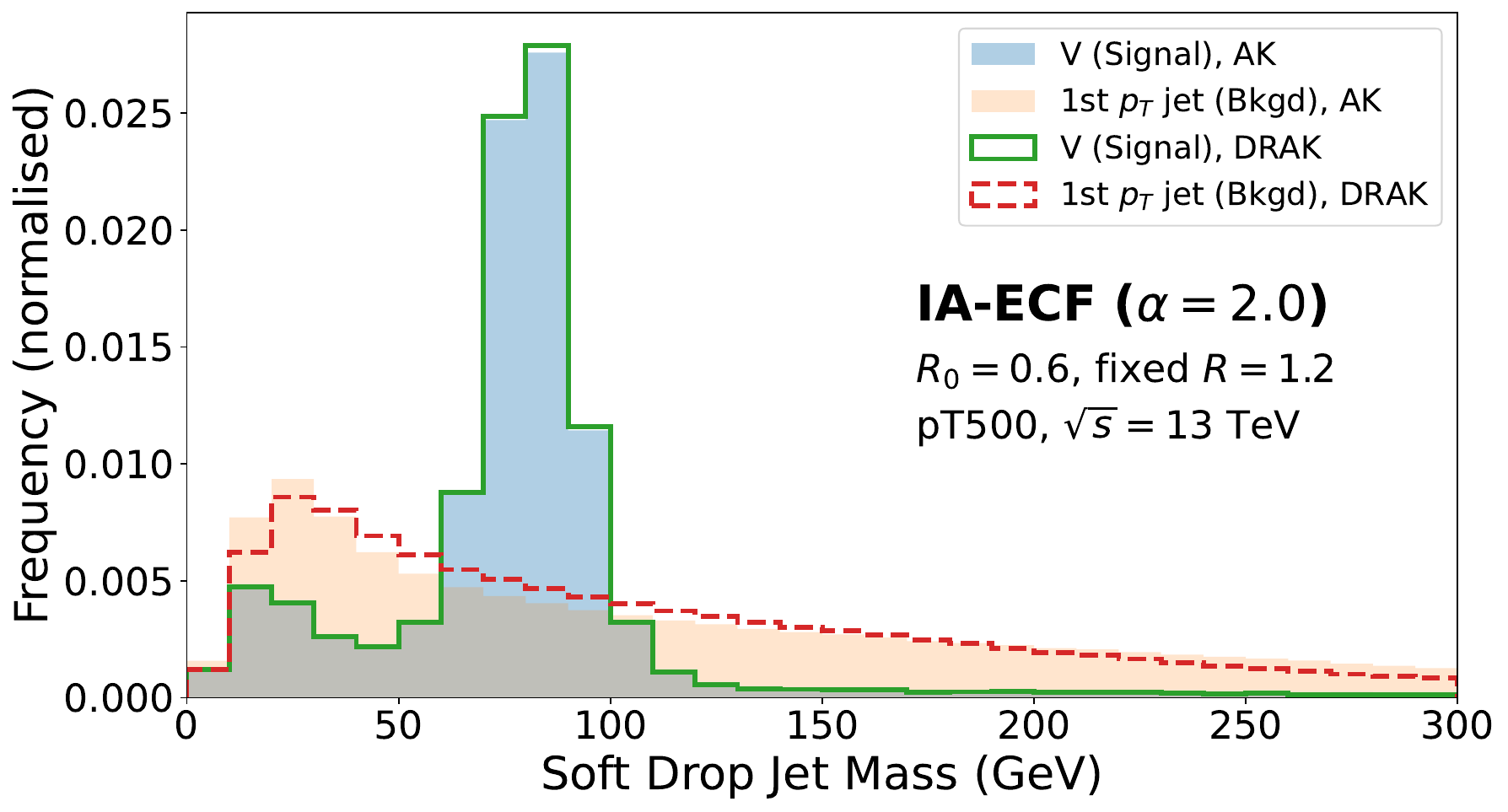}
\includegraphics[width=0.48\textwidth]{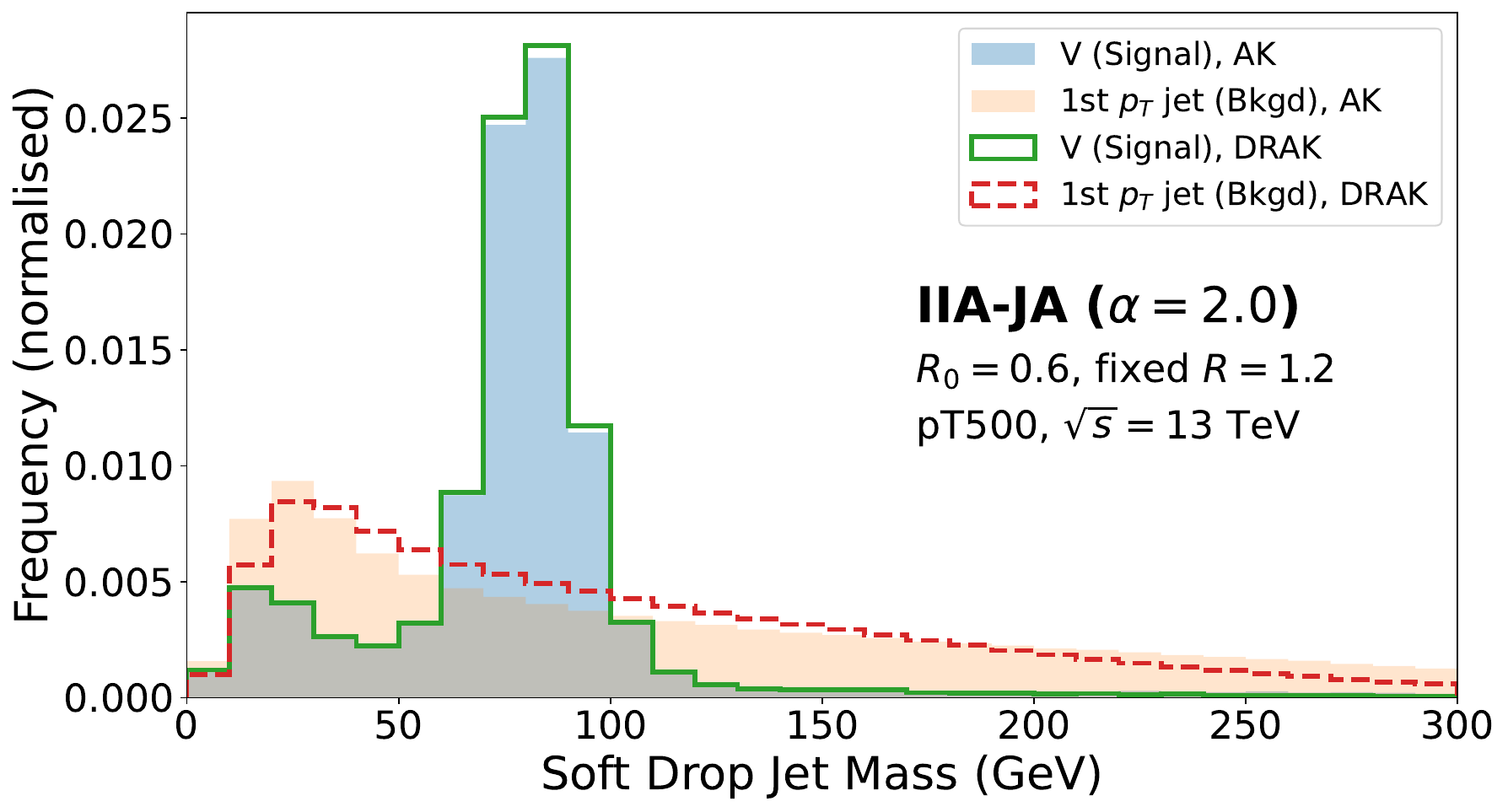}
\includegraphics[width=0.48\textwidth]{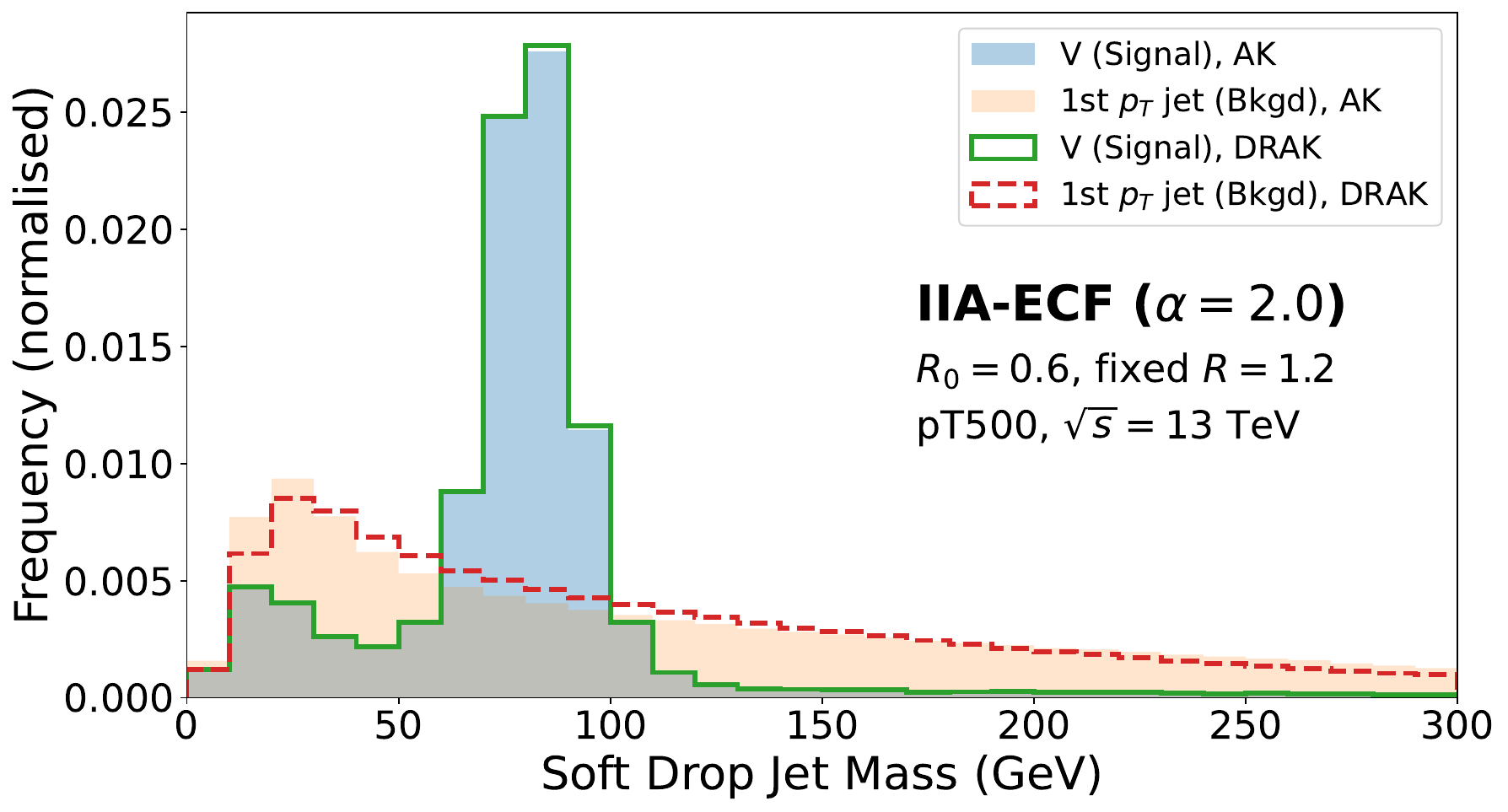}
\end{center}
\vspace{-16pt}
\caption{Normalised distributions of the Soft Drop jet mass for different DR variants for the $\Vjfh$ (signal) and $\jjfh$ (background) samples with starting radius $R_0=0.6$. The other conventions are kept the same as Figure~\ref{fig:Mass-Vj300-jj}. Here, we take $R=1.2$ for the fixed radius anti-$k_t$ algorithm.}
\label{fig:Mass-Vj500-jj}
\vspace{16pt}
\end{figure}

\begin{figure}[!h]
\begin{center}
\includegraphics[width=0.48\textwidth]{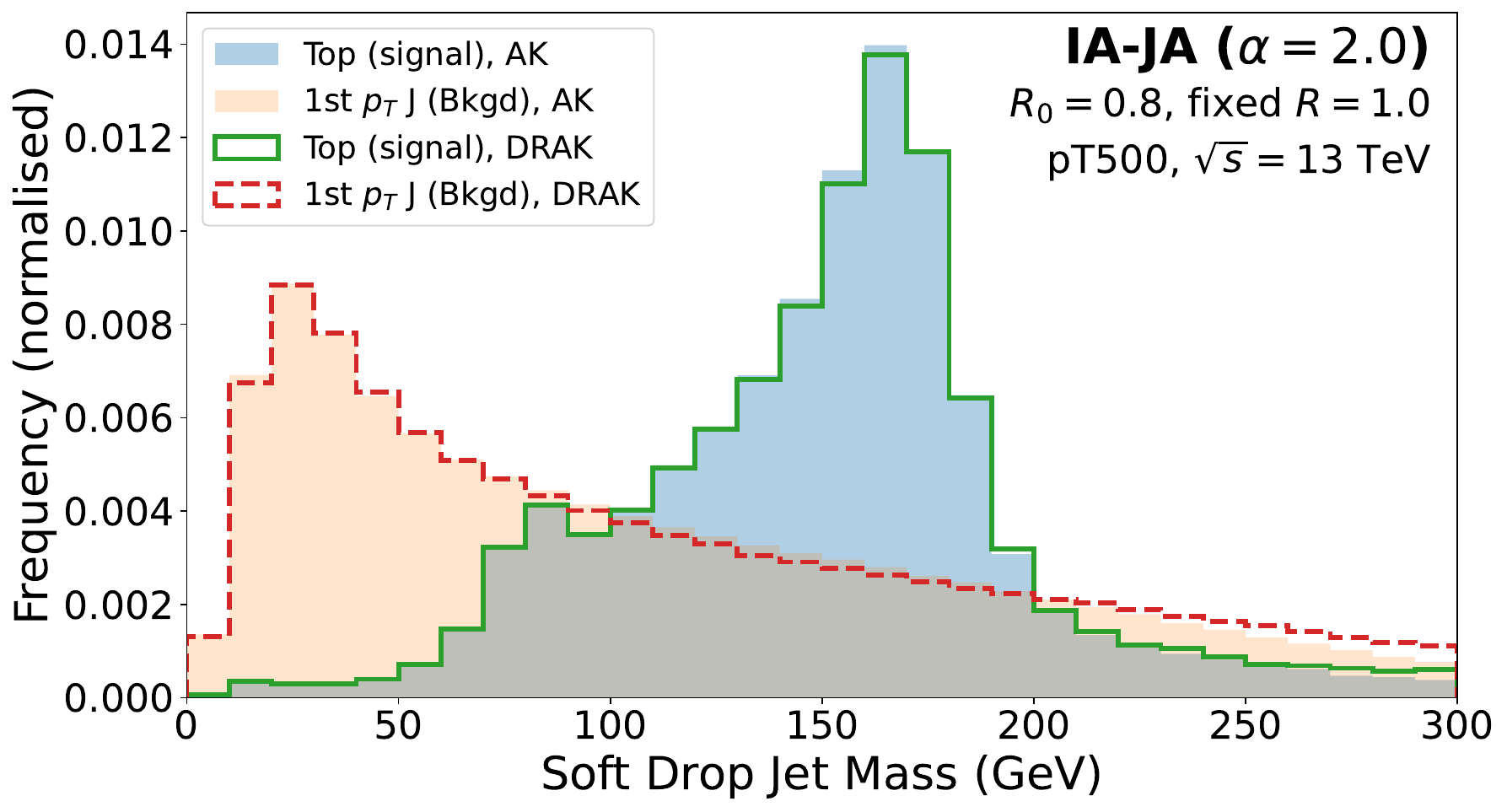}
\includegraphics[width=0.48\textwidth]{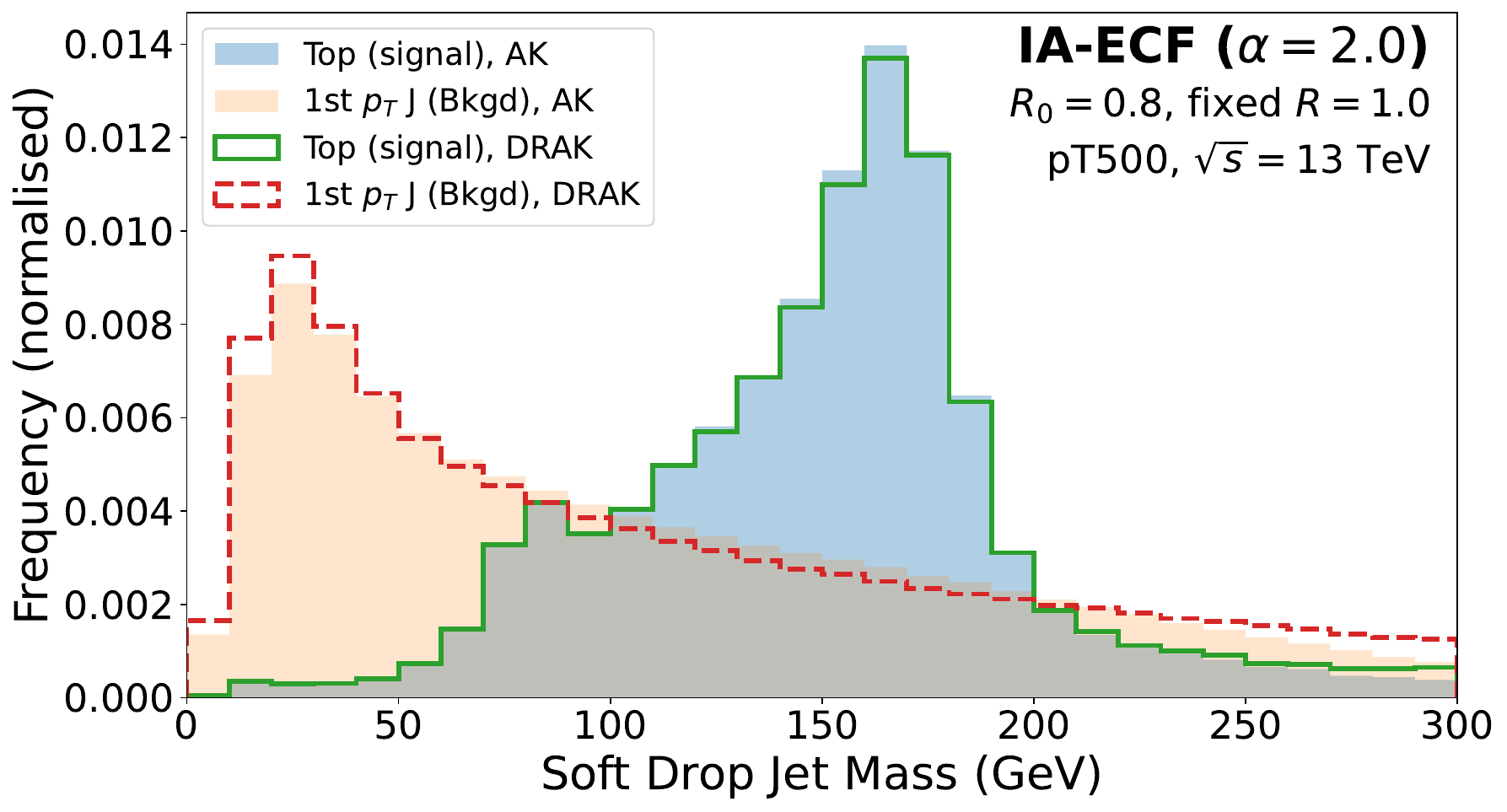}
\includegraphics[width=0.48\textwidth]{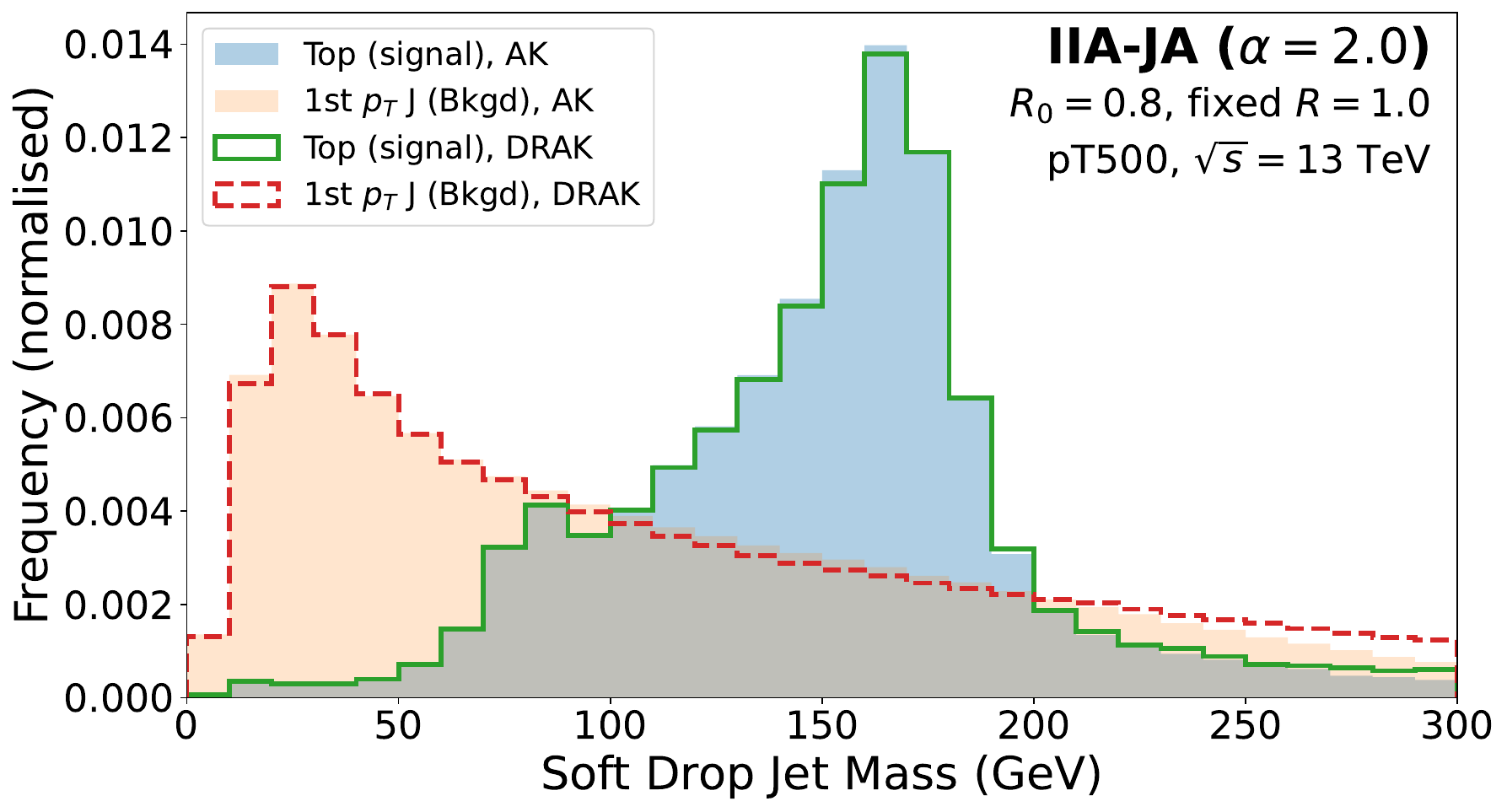}
\includegraphics[width=0.48\textwidth]{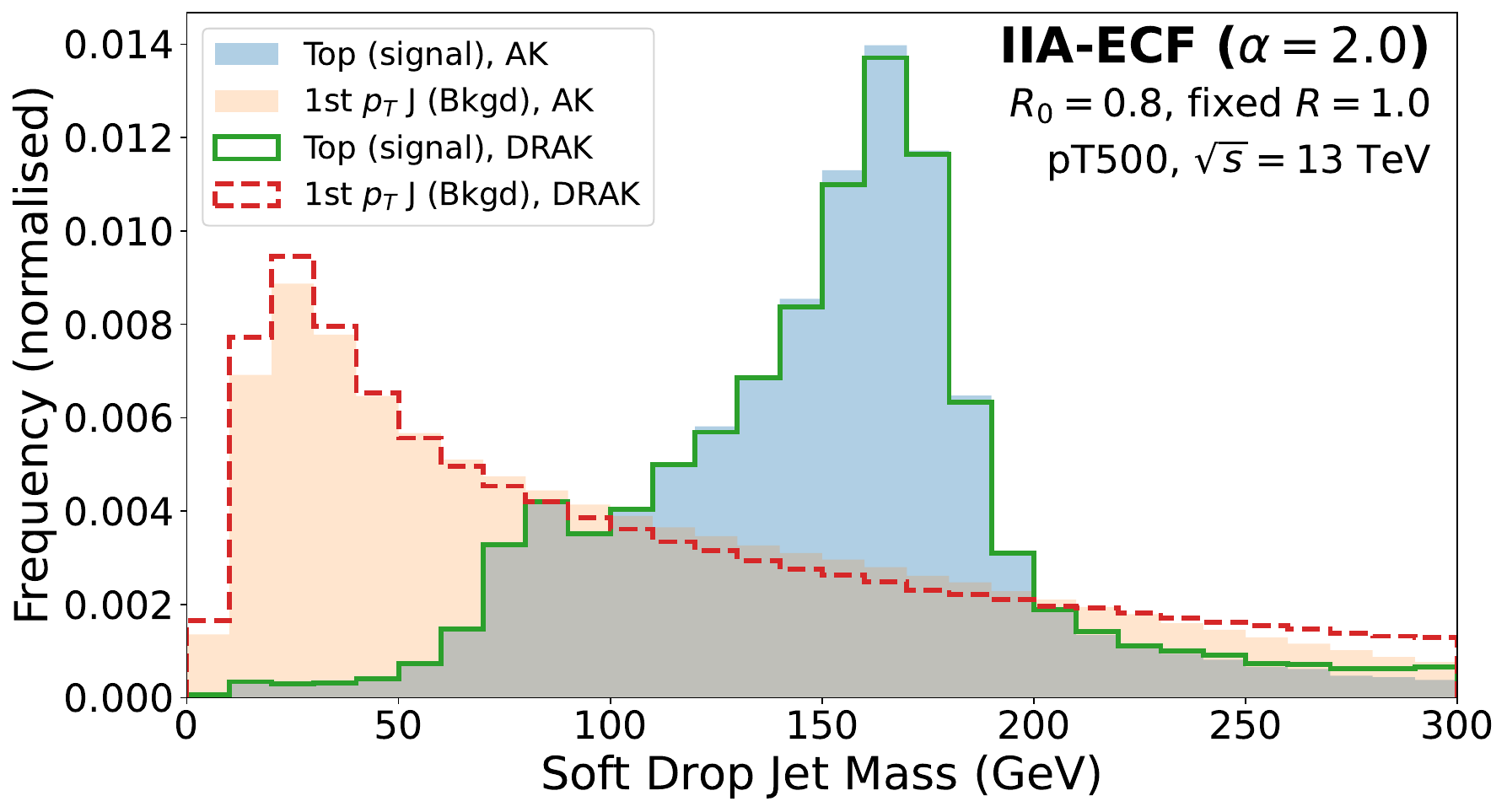}
\end{center}
\vspace{-16pt}
\caption{Normalised distributions of the Soft Drop jet mass for different DR variants for the $\tjfh$ (signal) and $\jjfh$ (background) samples with starting radius $R_0=0.8$. The other conventions are kept the same as Figure~\ref{fig:Mass-Vj300-jj}. Here, we take $R=1.0$ for the fixed radius anti-$k_t$ algorithm.}
\label{fig:Mass-tj-jj}
\end{figure}

The Soft Drop jet mass distributions for the different DR variants, along with distributions corresponding to the fixed-radius anti-$k_t$ algorithm, are shown in Figures~\ref{fig:Mass-Vj300-jj}, \ref{fig:Mass-Vj500-jj}, and \ref{fig:Mass-tj-jj} for the various signal and background samples.

\providecommand{\href}[2]{#2}\begingroup\raggedright\endgroup

\end{document}